\documentclass[aps,pre,twocolumn,superscriptaddress,showpacs,showkeys ]{revtex4-1}

\usepackage{graphicx}                   
\usepackage{hyperref}                   
\usepackage{amsmath,amssymb}            
\usepackage{epstopdf}
\usepackage{subcaption}					

\usepackage{color}
\definecolor{orange}{rgb}{1,0.5,0}
\definecolor{brown}{rgb}{0.65, 0.16, 0.16}
\definecolor{phlox}{rgb}{0.87, 0.0, 1.0}

\begin{document}

\title{Some Properties of Sandpile Models as Prototype of Self-Organized Critical Systems}

\author{M. N. Najafi}
\affiliation{Department of Physics, University of Mohaghegh Ardabili, P.O. Box 179, Ardabil, Iran}
\email{morteza.nattagh@gmail.com}

\author{S. Tizdast}
\affiliation{Department of Physics, University of Mohaghegh Ardabili, P.O. Box 179, Ardabil, Iran}
\email{Susan.tizdast@gmail.com}

\author{J. Cheraghalizadeh}
\affiliation{Department of Physics, University of Mohaghegh Ardabili, P.O. Box 179, Ardabil, Iran}
\email{jafarcheraghalizadeh@gmail.com }

\begin{abstract}
This paper is devoted to the recent advances in self-organized criticality (SOC), and the concepts. The paper contains three parts; in the first part we present some examples of SOC systems, in the second part we add some comments concerning its relation to logarithmic conformal field theory, and in the third part we report on the application of SOC concepts to various systems ranging from cumulus clouds to 2D electron gases. 
\end{abstract}


\maketitle

\section{Introduction}

Since the fascinating work of Bak, Tang and Wiesenfeld (BTW) coined the self-organized criticality (SOC) in $1987$~\cite{BTW1988Self} a huge number of papers appeared to explore various aspects of this term. It forms now a large class of critical phenomena. These systems show critical properties without tunning of any external parameter, for which the BTW sandpile model was the first prototype. Dhar discovered for the first time the Abelian structure of sandpiles so that we call them abelian sandpile models (ASM)~\cite{Dhar1990Self}. Despite its simple dynamics, ASM has various interesting features and numerous works, analytical and computational, have been done on this model~\cite{Majumdar1991Height,dDhar2006theoretical,Ivashkevich1994Waves,Dhar1994Inverse,Ktitarev1998Expansion,Majumdar1992Equivalence,Mahieu2001Scaling,Saleur1987Exact,Coniglio1989Fractal}. Thanks to conformal field theory (CFT), it is known that the BTW model is described with $c=-2$ class, $c$ being the central charge~\cite{dDhar2006theoretical}. Additionally the geometrical aspects of this model are understood in terms of its relation to loop-erased random walks (LERW)~\cite{Majumdar1992Exact}, which itself is related to Schramm-Loewner evolution (SLE) with the diffusivity parameter $\kappa=2$~\cite{Schramm2000Scaling,Najafi2012Avalanche}.\\

In this paper, after introducing the original work of BTW, we explore some properties of the model, with an emphasis on the dynamics of the avalanches in sandpiles. Its relation to Logarithmic Conformal Field Theory (LCFT) is explored. In the last part of the paper, we review various aspects of SOC in various systems, including the SOC in the fluid propagation in porous media, in cumulus clouds, in excitable complex networks, and in imperfect supports. We also describe the SOC techniques to explain the $1/f$ noise and metal insulator transition in two-dimensional electron gas (2DEG). We introduce also some generalizations of BTW, which are  the invasion BTW model, and the BTW model on vibrating systems. \\
The paper is organized as follows: in the following section, we introduce shortly the SOC in Nature and explain briefly the examples and avalanche dynamics and the basic ingredients. Section~\ref{BTW:Definition} is devoted to the definition of the BTW model and other variants. We will present relation to the logarithmic conformal field theory, including the ghost-free fields and W-Algebra in the Sec.~\ref{SEC:LCFT}. Various applications of SOC concepts to natural processes is presented in Sec.~\ref{SEC:application}.

\section{SOC in Nature}\label{SEC:Nature}
In nature, there are circumstances that unlike the thermal critical systems, no external parameter needs to be adjusted in order to reach and maintain in criticality, which are called self-organized critical (SOC) systems. Such a system automatically reaches and organizes itself in a critical state~\cite{markovic2014power}. These systems, which are usually open and absorb and dissipate energy, change with time but their general properties are almost unchanged on observed time scales. They need external energy to compensate the dissipation. \\

The original aim of the BTW model (to be explained in the next section) was to explain the "$1/f$ noise" phenomena that is seen in many natural systems, like rain fall~\cite{peters2001complexity}, sun flares~\cite{charbonneau2001avalanche,karakatsanis2012universality}, real piles of rice and other objects~\cite{munoz2001sandpiles,dickman2001critical}, earthquake~\cite{sornette1989self,telesca2001intermittent,gutenberg1936magnitude,davis1994multifractal,carlson1994dynamics,olami1992self,bak1989earthquakes,huang1998precursors}, forest fire~\cite{turcotte2004landslides}, and clouds~\cite{lohmann2016introduction,lovejoy1990multifractals,hentschel1984relative,cahalan1989fractal,joseph1990nearest}. The aim of this section is to introduce these natural phenomena as a motivation for analyzing the models of SOC.

\subsection{Examples}~\label{SEC:Examples}

\subsubsection{SOC in Earthquake}~\label{SEC:Earthquake}
One of the most popular examples of the SOC systems is earthquake for which the frequency of earthquakes ($ N $) with energy ($ E $) follows Gutenberg-Richter's power-law relation as follows~\cite{gutenberg1942earthquake,malamud2004landslide}
\begin{equation}
N = a{E^{ - b}}
\end{equation}
where $ N $ is the number of earthquakes with energy $ E $, $ a $ is the constant number that is a measure of the size and the amount of vibrational activity in the area, and $ b $ is a critical exponent being often between $ 0.8 $ and $ 1.2 $~\cite{sornette1992mean}. Also, the number of earthquakes is related to area $ A $ as follows
\begin{equation}
N = c{A^{ - d}}
\end{equation}
where $ d\approx 2.40 $~\cite{turcotte2004landslides} is another exponent. SOC in the earthquake is reported in many cases~\cite{sornette1989self}. Many SOC models have been developed to capture the physics of earthquakes, like block-spring models~\cite{sornette1992mean,nussbaum1987two}, sandpile based models~\cite{bak1989earthquakes,sornette1989self,olami1992self}, and sandpile on earthquake network~\cite{najafi2020avalanches}. In these models, the dynamics are predicted to be avalanche-like, based on a local stimulation (by increasing the local stress and tension), and the spread of stress throughout the system.\\

Generally, two strategies are often taken for explaining the observations of earthquakes: the quenched-disorder based models ascribing the observations of the seismic activities to the geometric and material irregularities in the earth, and the dynamical-instability models attributing the complexities to the stochastic forcing arising from the dynamic nonuniformities~\cite{carlson1994dynamics}. In the former, the power-laws observed in an earthquake is related to geometric features of the fault structure~\cite{kagan1987statistical}. Whether the earth is operating according to one of these schemes or in a hybrid one remains an open and fundamental problem. The application of SOC ideas for earthquake is very efficient and provides realistic results~\cite{bak1989earthquakes,sornette1989self,olami1992self}. The Olami-Feder-Christensen earthquake model model~\cite{olami1992self} is a two-dimensional coupled map lattice model which is known as a simplified version of the Burridge-Knopoff spring-block model~\cite{burridge1967model} for earthquakes. This model is famous and attracted much attention for it serves as a paradigm for nonconservative SOC systems, and also reproducing the most important statistical property of real earthquakes~\cite{olami1992self,olami1992temporal}, and also Omori's law~\cite{omori1894after,hergarten2002foreshocks}, and the statistics of foreshocks and aftershocks~\cite{helmstetter2004properties}.\\

In a recent study, the ideas of SOC models were applied to the Rigan earthquake~\cite{najafi2020avalanches}, in which a close relationship was observed between the dynamics of the SOC model and the real data of the earthquake. This relation can be understood by focusing on the stress propagation due to the tectonic motion of the continental plates, which is a slow steady process (like the other ``slowly driven'' SOC systems), but the release of stress occurs sporadically in bursts of various sizes. 

\subsubsection{SOC in Forest Fire}~\label{SEC:FFM}
Forest fire is another natural phenomenon that shows SOC behaviors, like power-law and scaling behaviors~\cite{clar1996forest,malamud1998forest}. Various surveys of firefighting data in different parts of the United States and Australia have shown a range of the size critical exponents between $ 1.3 $ and $ 1.5 $, depending on the area~\cite{malamud1998forest,turcotte2004landslides}. The SOC structure of forest fire was discovered by Malamud \textit{et. al.}~\cite{malamud1998forest}. Many models have emerged in order to capture the physics of this phenomenon. Among them, an important one is the Drossel-Schwabl model~\cite{drossel1992self,drossel1993forest} which in some limit gives acceptable exponents.\\

As a model for forest fire, let us consider the Drossel-Schwabl model~\cite{drossel1992self,drossel1993forest,grassberger2002critical,grassberger1993self}, defined on a $d$-dimensional lattice with lattice length $L$. In each time, each site of the system is empty or occupied by a green tree or a burning tree. At the initial time $t=0$ we suppose that the lattice sites are either occupied by green, or empty. The lattice state is updated at any time by the following rules:\\
\\
\textbf{1}. The burning site will be vacated in the next time,\\
\textbf{2}. The site where the green tree is located will catch fire at the next time if at least one of its nearest neighbor is burning, otherwise, it fires spontaneously with lightning probability $f$,\\
\textbf{3}. In an empty site, a tree grows with a $p$ probability.\\

Starting from an initial tree configuration, one The model can become critical only in the limit $p\rightarrow 0$ and $f/p\rightarrow 0$. In this limit, the length of the correlation is divergent~\cite{krenn2012natural}. The latter provides the conditions under which the time scale of tree growth and burning out the forest clusters are well-separated. The scale of the average size of clusters is controlled by the growth rate $\theta\equiv \frac{p}{f}$. In the simulations, at each time step $\theta$ sites are randomly chosen for being occupied (if it is already occupied nothing happens, otherwise it turns to occupied). Then a randomly chosen site is ignited so that all sites in the connected cluster to which the start site belongs burn. This model becomes critical in the limit $\theta\rightarrow \infty$, although there has been much discussion whether this model is SOC or not~\cite{grassberger2002critical}.

\subsubsection{SOC in Sun Flares}~\label{SEC:SF}
After the observation of R. C. Carrington and R. Hodgson in 1859 on the solar flares in white light, much attention has been paid to this problem. The cause of solar activities is the presence of a solar magnetic field in the hot plasma around the outer layer of the sun or the convective zone created by the collision of particles. The convective zone is the highest inner layer of the sun that extends from the radiative zone to the surface of the sun. This area is made up of convective effervescent cells. The magnetic flux forms active regions which include sunspot~\cite{ruzmaikin1990order,karakatsanis2012universality}.\\
		
Price et al.~\cite{karakatsanis2008soc} criticized the hypothesis of chaos in solar activity and found no evidence for a  low-dimensional deterministic nonlinear process using analysis of the sunspot number time series. Continuing this critical trend, the theory of Self- Organized Criticality (SOC) was proposed as the basic mechanism for explaining solar activity~\cite{baknature,vlahos2002esa}. According to the concept of SOC theory, the solar corona operates in a self-organized critical state, while the solar flares constitute random avalanche events with a power-law profile like the earthquake process. L.P. Karakatsanis and G.P. Pavlos ~\cite{karakatsanis2008soc} presented new results that strongly support the concept of low dimensional chaos and SOC theory. In this study, they found a co-existence between the self-organized critical state according to the soc theory and low dimensional chaotic dynamics underlying to the solar activity.
These results are obtained by nonlinear analysis of the sunspot index. For the original signal, the largest Lyapunov exponents were found to be zero, while the correlation integral slope profile was similar to the alternative data slope profile, showing a high-dimensional stochastic process and a critical state based on SOC theory ~\cite{bak1987self}.

\subsubsection{SOC in Rain-Falls}~\label{SEC:RF}
Another natural example of the SOC phenomenon is the rainfall, which is witnessed by the power-law behavior for the number of rain events versus size and the number of droughts versus duration~\cite{hurst1957suggested,peters2001complexity}. In this case, stable stimulation is provided by the heat of the sun, which causes the oceans to evaporate, for which rain relaxation with the continuous and uninterrupted event, rain. Peters \textit{et. al} showed that the accumulated water column displays scale-less fluctuations and also the number density of rain events per year $N(M)$ versus event size $M$ behaves like power-law, with exponent $1.36$, and the number density of droughts per year $N(D)$ versus drought duration $D$ with an exponent $1.42$. To understand the other quantity that was shown to be in power-law, let us define the rain-fall rate $q(u)\equiv \sum\limits_i {{n_i}} {V_i}{u_i} $, where $ {n_i} $ is the the density number of droplets with a volume of $ {V_i} $ that reaches the earth at the speed of $ {u_i} $. Then it was shown that 
\begin{equation}
R(\tau)/S(\tau)\sim \tau^H,
\end{equation}
where 
\begin{equation}
R(\tau)\equiv \text{max}_{1\leq t\leq\tau}X(t,\tau)-\text{min}_{1\leq t\leq\tau}X(t,\tau)
\end{equation}
and $X(t,\tau)\equiv\sum_{u=1}^t\left(q(u)-\left\langle q\right\rangle_{\tau}  \right) $, and $\left\langle q\right\rangle_{\tau}\equiv \frac{1}{\tau}\sum_{t=1}^{\tau}q(t)\Delta t$~\cite{peters2001complexity}. Self-organized criticality in rainfalls was observed in many other studies~\cite{andrade1998analysis,wang2012self,sarkar2006analysis,nnaji2011time,bove2006complexity}, which stimulated many theoretical studies on the subject~\cite{garcia2008applying,deluca2015data,pinho1998abelian,andrade2002new,andrade2003exact}.

\subsubsection{SOC in Clouds}~\label{SEC:Clouds0}
Self-affinity and scaling properties in clouds have been found from satellite images~\cite{Lovejoy}, and in particular in cumulus clouds~\cite{Austin} on several scales. Various observables were shown to exhibit scaling behavior, like the area-perimeter relation~\cite{Lovejoy,Chatterjee,Savigny,Austin,Malinowski,Batista,Madhushani}, the nearest neighbor spacing~\cite{Joseph}, the rainfall time series~\cite{Olsson}, cloud droplets~\cite{Malinowski2}, and the distribution function of geometrical quantities~\cite{Benner,Rodts,Yano,Gotoh}.
After these observations, and considering the multi-fractality of clouds~\cite{Lovejoy,Lovejoy2,Lovejoy3,Cahalan,Gabriel,Austin,Malinowski}, attempts for classifying clouds into universality classes were carried out based on cloud field statistics~\cite{Lovejoy4,Tessier,Pelletier} and cloud morphology~\cite{Sengupta}. The self-organized criticality in atmospher was first detected by Peters by analysing the precipitation~\cite{peters2006critical}, and developed further for atmospheric convective organization~\cite{yano2012self}. The fractal dimension of perimeter of self-organized vorties shown to be near $ \frac{4}{3} $ using the quasi geostrophic vorticity equation. The areaperimeter relation with the $ D = 1.37 \pm 0.02 $ of cirrus, and $ D = 1.18 \pm 0.05 $ for cumulunimbus tropical clouds. These fractal dimensions are in agreement with the relative turbulent diffusion model, predicting $1.35$, which is also confirmed by means of some other observations.\\

As a main building block of the atmosphere dynamics, turbulence seems to be essential in the dynamics and formation of clouds. Analysis of the images from the fair weather cumulus clouds reveals that they additionally exhibit self-organized criticality degrees of freedom, leading us to use the term SOC turbulent state. Observations (in our submitted paper) support the fact that this system, when projected to 2D, demonstrates conformal symmetry compatible with $ c=-2 $ conformal field theory, in contrast to 2D turbulence which is $ c=0 $ conformal field theory. Using a mix of turbulence and cellular automata, namely, the coupled map lattice model~\cite{Miyazaki2001A}, one obtains the same exponents as the observations. Also, in a separate (unpublished yet) work we developed a 2D monte carlo based stochastic model including the competition between avalanche dynamics and cohesive energy between water droplets that generates the same properties. The fractal geometry of clouds was seen in many real observations, like the multi-fractality structure of clouds, universality classes of cloud fields, analysis rainfall time series, nearest-neighbor spacing statistics, cumulus cloud morphology, "variable" and "steady" cloudy regions for warm continental cumulus cloud, fractal analysis of high-resolution cloud droplet measurements, the fractal dimension of noctilucent clouds, the fractal dimension of convective clouds around Delhi, scaling properties of clouds, self-similarity of clouds in the intertropical convergence zone, depending on the equivalent black body temperature.
\subsubsection{SOC in real piles}~\label{SEC:RealPiles}
SOC has been observed in labs for real piles, like pile of beads~\cite{ramos2009avalanche}, rice pile~\cite{amaral1996self,aegerter2003avalanche,frette1996avalanche,christensen1996tracer,amaral1996energy}, and other granular piles with various aspect ratio~\cite{denisov2012relation}. It has been observed that for dry sandpiles, macroscopic behaviors can be determined from the angle ${\theta_c}$~\cite{dDhar2006theoretical}. This angel is called the threshold angle, which depends on the structural details of its constituent grains. A sandpile with a local slope less than ${\theta _c}$ anywhere is stable, and adding a small amount of sand will cause a small reaction, but if this operation results in a slope larger than ${\theta _c}$, then an avalanche is formed, which sometimes is of the system size. On a pile with an average slope a bit less than ${\theta_c}$, the pile's response to the addition of sand is not very predictable. One possibility is that there will be no relaxation, or it may cause a medium-size avalanche or a catastrophic avalanche that will affect the entire system, called the critical state. Bak \textit{et. al.}~\cite{bak1987self} observed that if we build a pile by slowly pouring sand on a flat circular plate, we would get a conical pile with a slope equal to ${\theta_c}$. This system is constantly moving towards its critical state, this is an indication of SOC. The steady states of this process are described by the following properties:\\
Sand is being added to the system at a constant small rate, but it leaves the system in a very irregular manner, with long periods of apparent inactivity interspersed by events that may vary in size and which occur at unpredictable intervals. The interesting thing about these systems is that in steady-state, where the average amount of input and output energy is equal, the system exhibits critical properties. For example, power-law are found for various observations in the system, and the length of the correlation is infinite. Bak, Tang, and Wiesenfeld provided the BTW model that is an automatic cellular model for sandpile. This model is defined on the finite lattice. There is a positive integer variable at each site of the lattice, called the height of sandpile at the site ($ {z_i} $) and we have the threshold height for our lattice that called critical height ($ {z_c} $, it is often $ 2d $ that $ d $ is lattice dimension.) At each time step a site is picked randomly, and it's height $ {z_i} $ is increased by unity ($ {z_i} \to {z_i} + 1 $). If $ {z_i} > {z_c} $, this site is unstable. It relaxes by toppling whereby four sand grains leave the site, and each of the four neighboring sites gets one grain. If there is any unstable site remaining, it is toppled too. This process is an avalanche. The system in the BTW model reaches a stable state after the transition from transient states, which are called recurrent state because they are likely to be repeated during the casual process.\\
The rice pile model is another example of self-organized criticality phenomena. The mechanism, in this case, is that we consider two pieces of glass with a specific diameter of $5mm$ and a distance of $16mm$ from each other. We pour rice with a certain length from the middle of these two pieces of glass unit the system is stable, then we add colored grains similar to rice to the system and follow their movement so that we record the time of their entry and exit into the system. By obtaining the difference between the two terms, we measure the distribution of the length of time, the grains have been in the system. The following equation is obtained by plotting the time distribution in terms of lattice length ($L$):
\begin{equation}
P(T,L) = {L^{ - \beta }}F(\frac{T}{{{L^\nu }}})
\end{equation}

That $\nu$ and $\beta$ are the critical exponents. The statistical distribution of quantities in self-organized criticality phenomena is the power distribution.\\

There are other examples of SOC systems, like river basins~\cite{coulthard2007quantifying,van2010self}, in air pollution~\cite{shi2009self}, in climate change~\cite{liu2014self}, in brain plastisity~\cite{de2006self}, in stock markets~\cite{bartolozzi2005self,stauffer1999self}, in magnetosphere~\cite{valdivia2005magnetosphere}, in midlatitude geomagnetic activity~\cite{wanliss2010understanding}, in magnetohydrodynamics~\cite{uritsky2010structures}, in Kardar-Parizi-Zhang growth model~\cite{szabo2002self} in Bean state in YBCO thin films~\cite{aegerter2004self}, in granular systems~\cite{boguna1997long}, and in much more systems~\cite{paczuski1996avalanche} which are out of scope of this paper.

\subsection{Avalanche Dynamics and the Basic Ingredients}~\label{SEC:Dynamics}
It is a common belief that avalanche dynamics are the underlying mechanism that is responsible for SOC behaviors. Although normal
diffusive transport and branching are believed to be very basic ingredients of SOC, it was shown in~\cite{Manna2} that the avalanches in SOC can be linear as branchless random walks. In this case, their scaling behavior is different from that of branched avalanches. Sandpile models were introduced by Bak, Tang, and Wiesenfeld~\cite{BTW1988Self} (BTW) as a prototypical example for a class of models that show self-organized criticality. These models show critical behavior without fine-tuning of any external parameter. BTW model includes avalanche-based dynamics in which the system is slowly stimulated, i.e. subjected to small external perturbations. Large events in these systems, which are the result of these small stimuli, occur less frequently and on a larger scale, i.e. the energy is gradually absorbed and is excreted out on a larger scale. An interesting feature in these systems is that in the steady-state, where the average amount of energy input and output is equal, the system exhibits critical properties, which is a SOC state. The Abelian structure of the sandpile model was first discovered by Dhar so that it was thereafter named as Abelian sandpile model (ASM)~\cite{carlson1994dynamics}. Despite its simplicity, ASM has various interesting features and numerous works, analytical and computational, have been done on this model. Among them one can mention different height and cluster probabilities~\cite{kagan1987statistical}, the and avalanche distribution~\cite{langmuir1948production}, and also its the connection to the other models like the spanning trees~\cite{Lubeck1997BTW}, the ghost model~\cite{Najafi2012Observation,najafi2012avalanch}, and the $q$-state Potts model~\cite{najafi2016water}. For a good review see~\cite{stommel1947entrainment,dDhar2006theoretical}. Moreover, some of these results are analyzed in light of conformal field theory description with central charge $c = -2$~\cite{Majumdar1992Equivalence,najafi2012avalanch,Najafi2012Observation}, and also Schramm-Loewner evolution with the diffusivity parameter $\kappa=2$~\cite{najafi2012avalanch}.

\section{The BTW model}\label{BTW:Definition}

Let us consider the BTW on a two-dimensional square $d-$dimensional lattice hypercubic lattice with a linear size $L$ and coordination number $z=2d$. To each site a height variable $h_i$ is assigned which takes its value from the set ${1,2,...,z}$. This height variable shows the number of sand grains in the underling site. The dynamics of this model is as follows: in each step, one grain of sand is added to a randomly chosen site $i$, i.e., $h_i \rightarrow h_i + 1$. If the resulting height becomes more than $z$ (i.e. becomes unstable), the site topples and looses $2d$ grains of sand, each of which is transferred to one of $2d$ neighbors of the toppled site. As a result, the neighboring sites may become unstable and topple, and in this way, a chain of topplings may happen in the system until the system reaches a state with no unstable site. The chain of topplings then is called an avalanche. If a boundary site topples, one or two grains of sand (for the sites in the corners of the lattice two grains, and for the other boundary sites one grain) will leave the system. After reaching a stable configuration (the avalanche is finished), the process is repeated starting from another random site for grain injection. The toppling rule for the site $i$ can also be written in the form $h_j\rightarrow h_j+\Delta_{ij}$, where
\begin{equation}
\Delta_{ij}=\left\lbrace \begin{matrix}
-z & \text{if} \ j=i\\
+1 & \text{if} \ j \ \text{and} \ i \ \text{are neighbors}\\
0 & \text{otherwise}
\end{matrix}\right. 
\end{equation}
which is discrete Laplacian operatore. 

\subsection{Transient v.s. Recurrent Configurations}~\label{SEC:Configurations}
Let us suppose that we start from a random height configuration. Then during the system evolution, many configurations come about, some of which are transient, meaning that they do not occur again, and some of which are recurrent. In fact, the primitive configurations are transient, during which the average height grows with time (let us define the time as the number of injections). This linear growth cannot definitely last infinitely, and the system saturates at some stage (becomes stationary), after which the average height becomes nearly constant, meaning that on average energy input and output are the same. Recurrent states live in this regime. The total number of recurrent configurations is $\det\Delta$. One of the important questions is how we can identify a configuration to be recurrent or transient. Fortunately, there are tests that do this for us. One of our tests is the lack of forbidden subconfigurations (FSC) which does not exist in a recurrent configuration. FSCs can be identified simply by the requirement that they can never be created by the addition of sand and relaxation, if not already present in the initial state~\cite{dDhar2006theoretical}. We introduce two following test for checking transient or recurrent configurations:\\
We use the instability test of boundary site for the given configuration as follows: we add sand to any boundary site of configuration and we allow the system to evolve until achieving stable configuration. If the new configuration is the same as the first configuration, then the original configuration has been recurrent. The Second test is called the burning test. This test is a dynamic one in which we burn sites one by one. First, we assume that all the sites are unburned. In the next step, we burn all the sites whose heights are higher than the number of their unburned neighbors. If all the sites are burned at the end of the burning process, the desired configuration is recurrent~\cite{dDhar2006theoretical,dhar1990self}. 
\subsection{Other Sandpiles}~\label{SEC:OtherSandpiles}
\subsubsection{Manna Model}~\label{SEC:Manna}
The BTW model is a representative member of the BTW universality class. A relevant question is how one can change the details of this model to change its universality class. Stochasticity is one candidate to do this, which was tested for the first time by Manna by introducing a two-state SOC system, known also as the Manna model~\cite{manna1991two}. It includes randomness in the toppling  rule, i.e. in a two-dimensional system if a site has more than one sand, it is unstable and topples. During a toppling, a direction is chosen randomly (each direction is chosen with the probability of $0.5$) and all of the grains are distributed in this direction (no sand grain is transferred to the other direction). This model is interpreted as a realization od a system with particles experiencing a local infinite repulsive force between each other.\\
Much attention has been paid to identify  whether the Manna model belongs to the BTW universality class or not~\cite{Asasi,kadanoff1989scaling,manna1991critical,zhang1989scaling,dhar1989exactly,ben1996universality,christensen1993sandpile,chessa1999universality,dickman2003avalanche}, which is still open. The most challenging problem to this end is the accurate determination of exponents (controlling the finite-size effects, controlling the noise etc.).

\subsubsection{Zhang Model}~\label{SEC:Zhang}
The Zhang model~\cite{Zhang,Lubeck} indicates the continuous state of the BTW model, and the height of each site is considered to be a real number. The dynamics governing this model is such that at any moment, a continuous and random value between zero and one is added to the site. If its height exceeds a critical value, then that site is unstable, and its value is distributed to nearby sites, and its grain content becomes zero. This continues until all sites become stable. This model does not have an Abelian property, because the amount transferred in each toppling to neighboring sites depends on the initial value~\cite{dDhar2006theoretical}.
\subsubsection{The General Abelian Sandpile Model}~\label{SEC:General}
In sandpiles, the Abelian property is that the order of topplings in an avalanche dose not matter, both the interchanged topplings reaching to the same configuration. Let us describe a general set up for the Abelian sandpiles. We consider the model on a graph with $ N $ sites labeled by integers $i= 1,2,...,N $. We express the height of each site $ i $ with $ {z_i} $, which is a positive and integer number, and assign a threshold height $ z_i^c $ for each sites. At each time step, a site is chosen randomly and one sand grain is added to it. We are also given an integer $ N \times N $ toppling matrix $ \Delta  $, and a set of $N$ integers $ \left\{ {{z_{i,c}}} \right\} $, $ i=1,2,...,N $. If for any site $i$, $ {z_i} > z_i^c $, then the site is unstable and it topples. If $ {z_i} > z_i^c $ then $ {z_j} \to {z_j} - {\Delta _{ij}} $, for every $j$. We may choose $ z_i^c = {\Delta _{ii}} $, for which the allowed values of ${z_i}$
in a stable configuration are $ 1,2,...,{\Delta _{ii}} $. Evidently the matrix $\Delta $ has to satisfy some conditions to ensure that the model is well behaved~\cite{dDhar2006theoretical}.\\
\textbf{1}. ${\Delta _{ii}}>0$ for ever $i$ (otherwise toppling is never terminated).\\
\textbf{2}. For every pair $ i \ne j $, ${\Delta _{ij}} \le 0 $. This condition is required to establish the Abelian property.\\
\textbf{3}. $ \sum\limits_j {{\Delta _{ij}} \ge 0} $ for every $ j $ (this condition states that sand is not generated in the toppling process).\\
\textbf{4}. There is at least one site $ i $ such that $ \sum\limits_j {{\Delta _{ij}} > 0} $, called dissipative sites.
\subsubsection{Oriented/Directional Abelian Sandpile Model}~\label{SEC:Directional}
If we define an Abelian sandpile model on a directional lattice (e.g. the tilted square lattice), then we have oriented the Abelian sandpile model. In this model, the movement of the sand is in a certain direction and the threshold height is commonly considered to be unity. By adding one sand in a random site, after which the number of sand grains reaches to two (becomes unstable), then sand moves randomly to one of the bottom sites~\cite{dDhar2006theoretical}. The exponents here depend on the direction of the propagation, i.e. the time direction (top to bottom), or space direction (left to right).

\section{Relation to the Logarithmic conformal field theory}\label{SEC:LCFT}

Using the number of recurrent states in sandpiles (which is $\det \Delta$ as stated above), a connection is established with the free ghost field. From the properties of the Grassmann algebra, one can easily shown that $\det \Delta=\int \prod_{i=1}^{L}\text{d}\theta_i\text{d}\bar{\theta}_i\exp\left[\int \text{d}^2z\partial \theta(z)\bar{\partial}\bar{\theta}(\bar{z}) \right] $, where $\theta_i$ and $\bar{\theta}_i$ are independent Grassmann variables. Therefore the connection to the free ghost field is established defined by the following action
\begin{equation}
S=\int \text{d}^2z\partial \theta(z)\bar{\partial}\bar{\theta}(\bar{z})= \frac{1}{{2\pi }}\int {{\varepsilon _{\alpha \beta }}} \partial {\theta ^\alpha }\bar \partial {\theta ^\beta },
\label{Eq:Action}
\end{equation}
where the pair of free grassmanian scalar fields are defined as $  {\theta ^\alpha } = (\theta ,\bar \theta ) $, and $\epsilon_{\alpha\beta}$ is the cononical symplectic form, $ {\varepsilon _{12}} =  + 1 $, and  $ {\varepsilon ^{\alpha \beta }} =  - {\varepsilon _{\alpha \beta }} $. Using the $q$-state Potts model in the limit $q\rightarrow 0$, Majumdar and Dhar showed that ASM is equivalent to $c=-2$ conformal field theory in the scaling limit~\cite{majumdar1992equivalenc}. Especially the energy-energy correlation in $q$-state Potts model decays with distance $r$ like $r^{-2x_T}$, where~\cite{janke2004geometrical}
\begin{equation}
x_T=\frac{1+y}{2-y}
\end{equation}
where $y\equiv \frac{2}{\pi}\cos^{-1}\left(\frac{1}{2}\sqrt{q} \right) $. For $q\rightarrow 0$ we have $x_T=2$, which is a known result for height-height correlation is sandpiles~\cite{majumdar1992equivalenc}. Other correlations and probabilities in sandpiles can be expressed in terms of the grassmann fields and theirs derivatives~\cite{Majumdar1991Height}.

\subsection{Ghost Free Fields and $W$-Algebra}~\label{SEC:WAlgebras}
In this section, we turn to the continuum limit of the sandpile model, which is a nonunitary field theory. It is shown in~\cite{majumdar1992equivalenc} that it is a $c=-2$ conformal field theory (CFT) which is logarithmic. The term \textit{logarithmic} is used here due to appearing logarithmic correlations in CFT, i.e. LCFT. In these theories, each primary filed has a logarithmic partner, which enters in the ordinary operator product expansions (OPE)~\cite{gurarie1993logarithmic}. It can be shown that this is equivalent to adding a nilpotent exponent to the conformal dimension of primary fields~\cite{moghimi2001logarithmic,moghimi2003use}. In this way, much of what we have seen about a common conformal field theory  can be easily generalized to LCFT. The existence of ghostly fields in a field theory means the existence of a state with a negative magnitude, or in the other words, the theory in question is nonunitary. In ordinary CFT, the OPE of the energy-momentum tensor and a primary field gives singular terms, generating the corresponding conformal family for which one uses the Virasoro algebra~\cite{francesco1996conformal}. In LCFTs, in addition to the primary field $\phi $, we have a logarithmic partner $\psi $, which satisfies the following OPE with the energy-momentum tensor $T$~\cite{gurarie1993logarithmic}
\begin{equation}
\begin{array}{l}
T(z)\phi (\omega ) = \frac{{h\phi (\omega )}}{{{{(z - \omega )}^2}}} + \frac{{\partial \phi (\omega )}}{{z - \omega }} + ...\\
T(z)\psi (\omega ) = \frac{{h\psi (\omega ) + \phi (\omega )}}{{{{(z - \omega )}^2}}} + \frac{{\partial \psi (\omega )}}{{z - \omega }} + ....
\end{array}
\end{equation}
where $h$ is the conformal dimension of $\phi$. The action of the Virasoro operator at the zero levels ${L_0}$ on the pair $\phi $ and $\psi $ has a jordan form, giving rise to logarithmic terms in the corresponding correlation functions. The infinitesimal transformation of the pair is

\begin{equation}
\begin{array}{l}
{\delta _\varepsilon }\phi (z) = (h{\partial _z}\varepsilon  + \varepsilon {\partial _z})\phi (z)\\
{\delta _\varepsilon }\psi (z) = (h{\partial _z}\varepsilon  + \varepsilon {\partial _z})\psi (z) + {\partial _z}\varepsilon \phi (z),
\end{array}
\end{equation}
The calculations become simpler if we use the mixed opertor $ \phi (z,\lambda ) = \phi (z) + \lambda \psi (z) $ where $\lambda$ is a nilpotent number ($ {\lambda ^2} = 0 $). Then the above equations cast to the following abbrivated form
\begin{equation}
{\delta _\varepsilon }\phi (z,\lambda ) = ((h + \lambda ){\partial _z}\varepsilon  + \varepsilon {\partial _z})\phi (z,\lambda ).
\end{equation}
using of which one obtains the finite transformation ($z\rightarrow \omega(z)$)
\begin{equation}
\phi (z,\lambda ) = {\left( {\frac{{\partial \omega }}{{\partial z}}} \right)^{h + \lambda }}\phi (\omega ,\lambda ).
\end{equation}
One can derive the two point function of the mixed fields that is invariant under the translation, scale, rotation and special conformal transformations
\begin{equation}
\left\langle {\phi ({z_1},{\lambda _1})\phi ({z_2},{\lambda _2})} \right\rangle  = \frac{{a({\lambda _1},{\lambda _2})}}{{{{(z - \omega )}^{2h + {\lambda _1} + {\lambda _2}}}}},
\end{equation}
where $ a({\lambda _1},{\lambda _2}) = {a_1}({\lambda _1} + {\lambda _2}) + {a_{12}}{\lambda _1}{\lambda _2} $, resulting to $ \left\langle {\phi ({z_1})\phi ({z_2})} \right\rangle  = 0 $. Similar results can be derived for otheer two point correlations, and also the three point functions
\begin{equation}
\left\langle {\phi ({z_1},{\lambda _1})\phi ({z_2},{\lambda _2})\phi ({z_3},{\lambda _3})} \right\rangle  =\frac{ f({\lambda _1},{\lambda _2},{\lambda _3})}{z_{12}^{ {a_{12}}}z_{23}^{ {a_{23}}}z_{31}^{ {a_{31}}}},
\end{equation}
where
\begin{equation}
\begin{split}
a_{ij} &= {h_i} + {h_j} - {h_k} + {\lambda _i} + {\lambda _j} - {\lambda _k}\\
f({\lambda _1},{\lambda _2},{\lambda _3}) &= \sum\limits_{i = 1}^3 {{c_i}} {\lambda _i} + \sum\limits_{1 \le i < j \le 3} {{c_{ij}}} {\lambda _i}{\lambda _j} + {c_{123}}{\lambda _1}{\lambda _2}{\lambda _3}
\end{split}
\end{equation}
and $ {z_{ij}} = ({z_i} - {z_j}) $ and $ k $ is the index other than $i$ and $j$, and $c_i$, $c_{ij}$ and $c_{123}$ are constants that cannot be determined using global conformal invariance.\\

Now let us turn to our problem. Using simple grassmann calculations one can easily show that the quantum expectation on unity $\left\langle \textbf{1}\right\rangle $ is zero. To understand this, let us expand the ghost fields in the Euclidian coordinate $z$
\begin{equation}
{\theta ^\alpha }(z) = \sum\limits_{n \ne 0} {\theta _n^\alpha } {z^{ - n}} + \theta _0^\alpha \log (z) + {\xi ^\alpha },
\end{equation}
where $\theta_n$ and $\bar{\theta}_n$ are modes. The absence of zero-mode $\xi$ and $\bar{\xi}$ in the action (Eq.\ref{Eq:Action}) leads to vanishing of the expectation value of unity. In the above equation, $ n $ is an integer number for the untwisted sector and it is a half-integer number for the twisted sector. It is appropriate to insert the zero modes $\xi\bar{\xi}$ in the expectations in order to avoid vanishing the correlation functions involving $ \theta $ fields. For example, the two-point correlation function is
\begin{equation}
\left\langle {{\theta ^\alpha }(z){\theta ^\beta }(\omega )\bar \xi \xi } \right\rangle  = {\varepsilon ^{\alpha \beta }}\log \left| {z - \omega } \right|
\end{equation}
The $\theta$ fields are not primary, but their derivative $\partial\theta$ and $\partial\bar{\theta}$ are:
\begin{equation}
\left\langle {\partial {\theta ^\alpha }(z)\partial {\theta ^\beta }(\omega )} \right\rangle  = {\varepsilon ^{\alpha \beta }}\frac{1}{{2{{(z - \omega )}^2}}}.
\end{equation}
The energy-momentum tesor is $ T = 2:\partial \theta \bar \partial \bar \theta : $ resulting to the following OPE with the central charge $c=-2$:
\begin{equation}
T(z)T(\omega ) = \frac{{ - 1}}{{{{(z - \omega )}^4}}} + \frac{{2T(\omega )}}{{{{(z - \omega )}^2}}} + \frac{{\partial T(\omega )}}{{z - \omega }} + ....
\end{equation}
One can also find the same central charge using the explicit for of Virasoro operators $L_n$ which is an expansion in terms of the modes
\begin{equation}
L_n=2\sum_m :a_ma_{n-m}:
\end{equation}
where
\begin{equation}
a_n=\left\lbrace \begin{matrix}
n\theta_n & n \ne 0\\
-\theta_0 & n=0
\end{matrix}\right. ,
\end{equation}
and $::$ means the normal ordering. One can easily check that $T(z)=\sum_n z^{-n-2}L_n$. One may try to build the Fock space using the modes and a vacuum state. One should however take into account that in this case, the Verma module is staggered, meaning that a descendant of the primary field may be itself a primary field having its own Verma module~\cite{kytola2009staggered}. 
There are three important representations $ \left( {{R_0},{R_1},R} \right) $ in $c=-2$ models, the two representations $ {R_0} $, $ {R_1} $ are the highest weight representations (explored in the following) but $ R $ is local representation whose amplitudes are local.
The $ {R_0} $ representation contains the identity operator $I$ and the field $ \tilde{I}\equiv -2:\theta \bar \theta : $ (with zero conformal dimension) whose OPE with $ T $ is shown to be
\begin{equation}
T(z)\tilde{I} (\omega ) = \frac{{ I}}{{{{(z - \omega )}^2}}} + \frac{{\partial \tilde{I}}}{{z - \omega }} + ...
\end{equation}
showing that $\tilde{I}$ is the logarithmic partner of $I$. The ${R_1}$ representation includes $ {\phi ^\alpha } = \partial {\theta ^\alpha } $ and $ {\psi ^\alpha } = :\partial {\theta ^\alpha }\tilde I: $ carrying the conformal weight $(1,0)$ and the OPE
\begin{equation}
T(z){\psi ^\alpha }(\omega ) = \frac{{{\theta ^\alpha }(\omega )}}{{2{{(z - \omega )}^3}}} + \frac{{{\phi ^\alpha }(\omega ) + {\psi ^\alpha }(\omega )}}{{{{(z - \omega )}^2}}} + \frac{{\partial {\psi ^\alpha }(\omega )}}{{z - \omega }} + ...
\end{equation}
which is anomalous OPE, having an extra singular term (the first term in the right hand side). This make $R_1$ representation more complex than $R_0$, and its operator content is larger. In fact $W$-algebras come about in this representation which works with operators at level three~\cite{gaberdiel2003algebraic}
\begin{equation}
\begin{array}{l}
{W^ + } = {\partial ^2}\theta \partial \theta \\
{W^0} = \frac{1}{2}({\partial ^2}\theta \partial \bar \theta  + {\partial ^2}\bar \theta \partial \theta )\\
{W^ - } = {\partial ^2}\bar \theta \partial \bar \theta .
\end{array}
\end{equation}
Note that this set is isospin one with spins $+1$, $0$ and $-1$ for first, second and third fields respectively (note that conformal weights are $(1,0), (1,1)$ and $0,1$ respectively)~\cite{Blumenhagen19991}. Their OPEs are
\begin{equation}
\begin{array}{l}
{W^i}(z){W^j}(\omega ) = {g^{ij}}(\frac{1}{{{{(z - \omega )}^6}}} - 3\frac{{T(\omega )}}{{{{(z - \omega )}^4}}} - \frac{3}{2}\frac{{\partial T(\omega )}}{{{{(z - \omega )}^3}}}\\
+ \frac{3}{2}\frac{{\partial {T^2}(\omega )}}{{{{(z - \omega )}^2}}} + 4\frac{{{T^2}(\omega )}}{{z - \omega }} + \frac{1}{6}\frac{{\partial {T^3}(\omega )}}{{z - \omega }} - 4\frac{{\partial {T^2}(\omega )}}{{z - \omega }})\\
- 5f_k^{ij}(\frac{{{W^k}(\omega )}}{{{{(z - \omega )}^3}}} + \frac{1}{2}\frac{{\partial {W^k}(\omega )}}{{{{(z - \omega )}^2}}} +
\frac{1}{{25}}\frac{{{\partial ^2}{W^k}(\omega )}}{{z - \omega }} + \frac{1}{{25}}\frac{{(T{W^k})(\omega )}}{{z - \omega }})
\end{array}
\end{equation}
Where $ {g^{ij}} $ is the metric on the isospin one representation, $ {g^{ +  - }} = {g^{ -  + }} = 2 $ and $ {g^{00}} =  - 1 $, and $ f_k^{ij} $ are the structure constants of SL(2). One can write the $W$ algebra, by using the above OPEs. Following Gaberdial and Kausch~\cite{gaberdiel2003algebraic}, we have:
\begin{equation}
\begin{array}{l}
\left[ {{L_m},W_n^i} \right] = (2m - n)W_{m + n}^i\\
\left[ {W_m^i,W_n^j} \right] = {g^{ij}}(2(m - n){\Lambda _{m + n}}\\
+ \frac{1}{{20}}(m - n)(2{m^2} + 2{n^2} - mn - 8){L_{m + n}}\\
- \frac{1}{{120}}m({m^2} - 1)({m^2} - 4){\delta _{m + n}})\\
+ f_k^{ij}(\frac{5}{{14}}(2{m^2} + 2{n^2} - 3mn - 4)W_{m + n}^k + \frac{{12}}{5}V_{m + n}^k)
\end{array}
\end{equation}
Where $ \Lambda  = :{T^2}: - \frac{3}{{10}}{\partial ^2}T $ and $ {V^a} = :T{W^a}: - \frac{3}{{14}}{\partial ^2}{W^a} $ are quasiprimary normal ordered fields. This $ W $ algebra is different from Zamolodchikov's $ W $ algebra~\cite{zamolodchikov1995infinite}, because $ f_k^{ij} $ is different. By using the above $ W $ algebra and Gaberdiel and Kausch~\cite{gaberdiel2003algebraic} found the null vectors, from which, using the zero mode the following equation for highest weight field $\phi$ were found~\cite{gaberdiel2003rationality,gaberdiel2003algebraic}: 
\begin{equation}
L_0^2(8{L_0} + 1)(8{L_0} - 3)({L_0} - 1)\phi  = 0.
\end{equation}
implying that $ h $ must be from the set $\left\lbrace  0,\frac{{ - 1}}{8},\frac{3}{8},1 \right\rbrace $ which are represented by $V_h$. For $V_0$ we have $L_0^2\phi=0$ which is satisfied for $I$ and $\tilde{I}$ that is the only logarithmic highest weight representation of $c=-2$, and the other logarithmic representations are not highest weight (see $R_1$ representation for example). Two of the three remaining highest weight representations are related to the twisted sector and the other is nontwisted. For a more complete reference see~\cite{rajabpour2007some}.

\section{Application of Sandpiles to Natural Processes}\label{SEC:application}
In this section, we turn to the application of SOC concepts to real systems. All subjects considered here are simulations (except a part for SOC in clouds). Most parts of this section are carried out by the authors of this paper.
\subsection{SOC in Fluid Propagation in Porous Media}~\label{SEC:PorousMedia}
\begin{figure*}
	\centerline{\includegraphics[scale=.35]{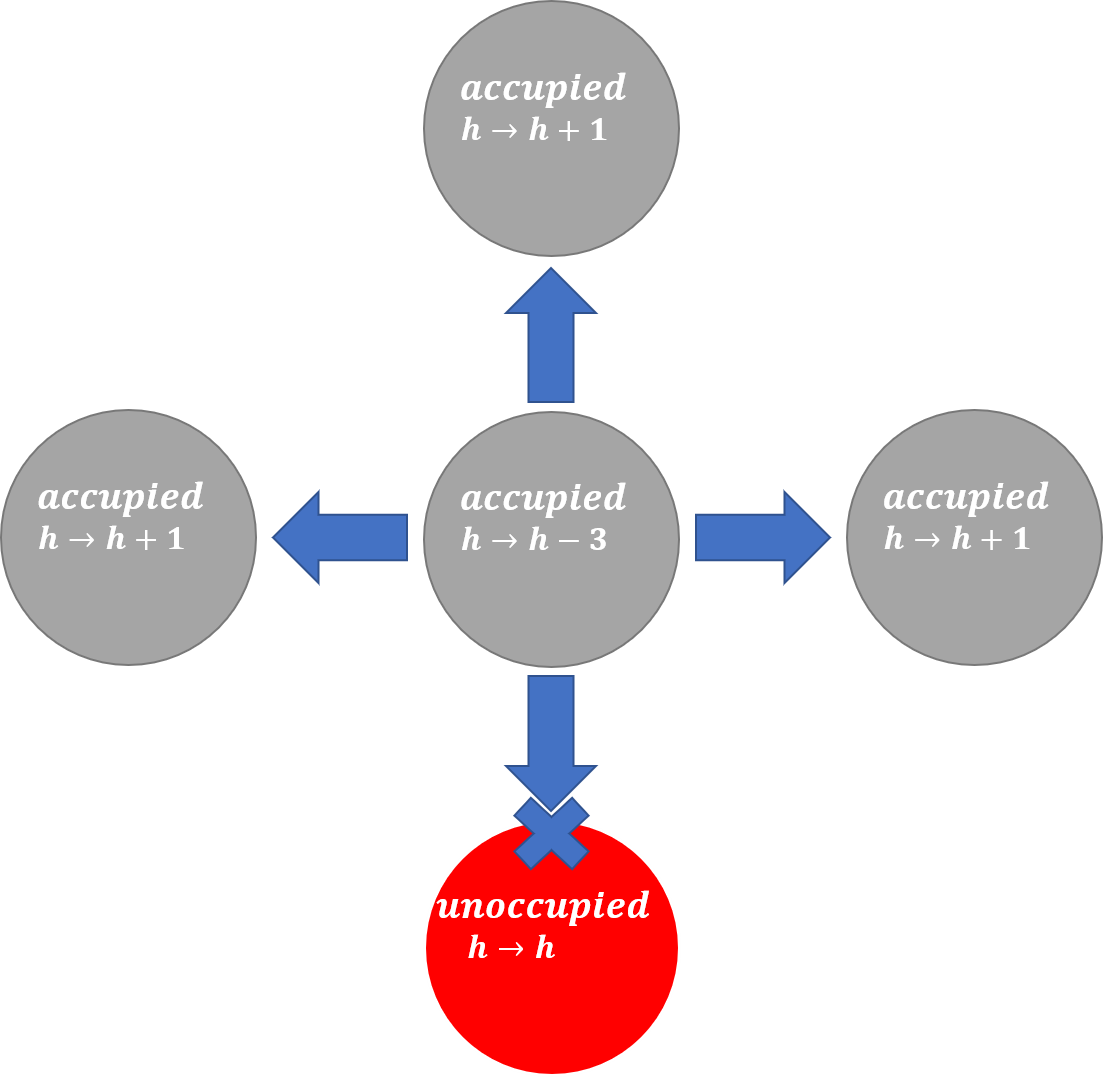}}
	\caption{SEC.~\ref{SEC:application}:(a) The toppling rule for a typical site which has three occupied neighbors }
	\label{Fig:BTW}
\end{figure*}
\begin{figure*}
	\centerline	{\includegraphics[scale=.09]{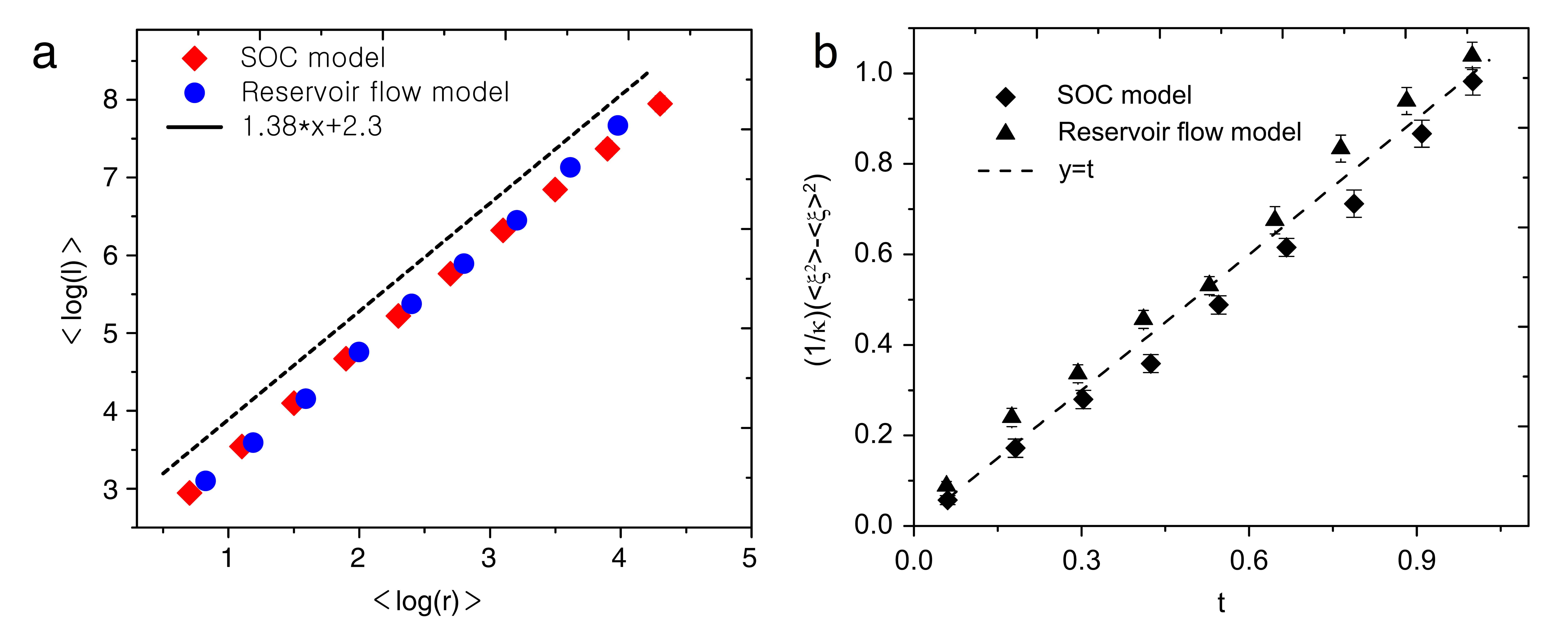}}
	\caption{SEC.~\ref{SEC:application} :(a) The plot of $\langle \log ( l )\rangle$  versus $\langle \log ( r ) \rangle$ (fractal dimension) for the reservoir flow model and SOC model. (b) $ \langle \xi^2 \rangle -\langle \xi \rangle^2$ versus $t$ for the case $p=p_c$ }
	\label{Fig:BTW_p}
\end{figure*}

The fluid propagation in porous media (involving avalanche-type dynamics) is a complex procedure, resulting in various interesting patterns. The avalanches arise from a non-linearity in the laws governing the dynamics of fluid in this system, namely the \textit{critical water saturation}. The fluid in a typical region of the reservoir is static (it does not macroscopically transfer to the neighboring regions), until the accumulated fluid saturation in that region exceeds a certain saturation known as critical saturation $S_C$ above which the fluid overflows freely (governed by the pressure gradient) to the neighboring regions. Therefore, for a region (consisting of many pores), the total water saturation is small, no fluid transfers to the other regions. Once the amount of fluid in that region reaches $S_C$, then some small droplets of fluid in the pores aggregate so that the fluid acquires the ability to move to the neighboring regions. In terms of the porous media parameters, one common choice for the relation between the relative permeability of phase $\alpha$ (shown by $k_{r\alpha}$) and the saturation of that phase ($S_{\alpha}$) is
\begin{equation}
k_{r\alpha}=\left\lbrace \begin{matrix}
S_{\alpha}-S_C & S_{\alpha}\geq S_C\\
0 & S_{\alpha}< S_C
\end{matrix}\right.
\end{equation}
which shows the above-mensioned dynamics. In~\cite{najafi2016water}, it was shown that the set of Darcy equations (known as the reservoir flow or RF model) for two-phase propagation in porous media is very similar to the BTW dynamics, except that the former is directional (the fluid moves in the direction of the pressure gradient), whereas the latter is not. In this work the ordinary BTW model has been used defined on a percolation lattice which realizes a porous media (uncorrelated) in which some points are occupied with a probability of $p$ and the others are unoccupied (with the probability $1-p$). Fig.(\ref{Fig:BTW}) shows the toppling rule for a typical site that has three occupied neighbors. To find the connection between this model and the Darcy model, the Schramm-Loewner evolution theory (SLE) theory was used and some other standard statistical analyses were performed such as fractal dimension on their domain-walls to investigate its behavior in terms of $p$. The results of the SLE theory are depicted in Fig.(\ref{Fig:BTW_p}) where $\xi_t$ , known as the driving function in the SLE theory is a continuous real-valued function that is shown to be proportional to the one-dimensional Brownian motion ( $\xi_t =\sqrt{\kappa}B_t$, $\kappa$ being the diffusivity parameter) for the conformal invariant curves \cite{cardy2005sle}. Two critical models with the same diffusivity parameter are believed to be in a same universality class. This graph shows that the RF and the SOC models are equivalent on the percolation threshold $p=p_c$, being compatible with the Ising universality class. \\

Evidently the permeable pores of the porous media are not completely independent in the natural systems. As an example, consider the sedimentation process of the reservoir rock in which some parts of the reservoir rock become impermeable to flow. Therefore, the correlation of the unoccupied sites depends on the dynamics of the sedimentation process. In \cite{cheraghalizadeh2017mapping} it was assumed that the correlation of the occupied sites is given by a zero-magnetic field, ferromagnetic Ising model. The strength of the correlations is controlled by the Ising coupling constant and the artificial temperature $T$.
This model shows that at the Ising critical temperature $T_c$, the model is compatible with the universality class of two-dimensional (2D) self-avoiding walk (SAW). The mixing of two conformal symmetric models is also of interest to the theoretical side. Mathematically this problem can be tracked in terms of the Zamolodchkiv’s $c$-theorem, in which knowing the scaling perturbing filed, one can obtain the change of central charge of the conformal field theory $\delta_c = c_{\text{IR}} - c_{\text{UV}}$ \cite{schramm2011scaling}. The effect of the second CFT model can presumably be codded in a scaling field from the operator content of the original (first) CFT model.\\
In another work \cite{PhysRevE.101.032116} Najafi \textit{et al.} concentrated on the SOC dynamics on 3D correlated porous media. It was found evidence for a new nonequilibrium universality class that is reached by changing the geometry of the underlying graph upon which the model is defined. This might be applicable to experiments with spatial flow patterns of transport in heterogeneous porous media~\cite{oswald1997observation}.

\subsection{SOC in Cumulus Clouds}~\label{SEC:Clouds}
\begin{figure*}
	\centerline{\includegraphics[scale=.14]{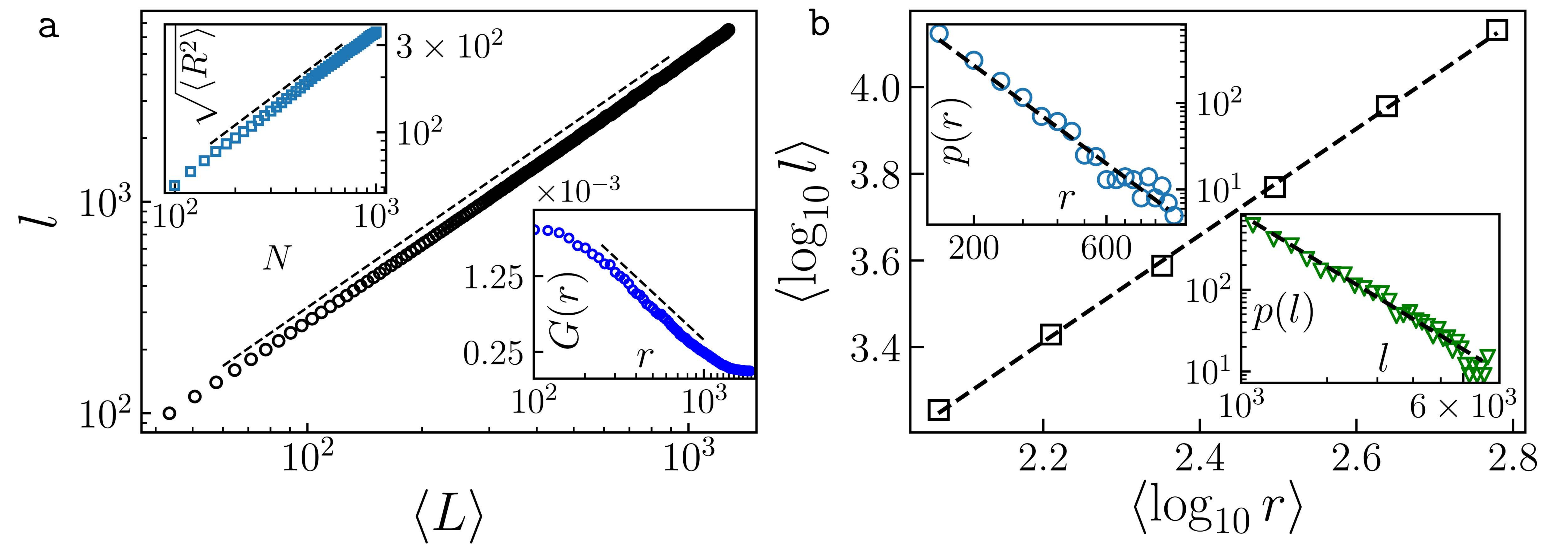}}
	\caption{SEC.~\ref{SEC:Clouds}:(a) The log-log plot of trace lengths $l$ in terms of $L$ (the box linear size). The dashed line is a linear fit with slope $D_f^a = 1.248 \pm 0.006$. Upper inset is the end-to-end distance $R$ in terms of $N$, and the lower inset is a semi-log plot of the loop green function in terms of $r$, with the exponent
		$\nu = 0.81 \pm 0.01$. (b) Ensemble average of $\log l$ in terms of $\log r$ ($\langle \rangle$ means the ensemble average) with slope $D_f^b = 1.22 \pm 0.02$. The log-log plot of the distribution function of $r$ and $l$ are shown in
		the upper and lower insets , with exponents $\tau_r = 2.12 \pm 0.03$ and $\tau_l = 2.38 \pm 0.02$ respectively. }
	\label{Fig:cloud}
\end{figure*}
As stated in the previous sections, there are many papers reporting on the fractal structure of the clouds, some of which are based on SOC. Convective clouds (such as cumulus clouds as a member of the cumuliform clouds) develop in unstable air due to the buoyancy force, resulting from water vapor, supercooled water droplets, or ice crystals, depending upon the ambient temperature. Cumulus clouds have flat bases and are often described as \textit{cotton-like} low-level clouds, less than 2 km in altitude unless they are more vertical (cumulus congestus form). These clouds which may appear in lines or in clusters (producing little or no precipitation) are often precursors of other types of clouds, such as cumulonimbus when influenced by weather factors such as instability, moisture, and temperature gradient. When they grow into the congestus or cumulonimbus clouds, they are more probable to precipitate. The height of the cloud (from its bottom to its top) depends on the temperature profile of the atmosphere. \\

Cumulus clouds form via atmospheric convection as air warmed by the surface begins to rise, resulting in the temperature decrease and humidity rise. At a threshold, named as lowest condensation level (LCL), in which the relative humidity reaches $100\%$, then condensation to the wet phase (known as wet-adiabatic phase) starts. The released latent heat (due to condensation) warms up the air parcel, resulting to further convection. At LCL, the nucleation process starts on various nuclei present in the air. The process of formation of raindrops and rainfall has been explained successfully by Langmuir~\cite{langmuir1948production}. \\
Although the liquid water density within a cumulus cloud changes with height above the cloud base~\cite{stommel1947entrainment} (for the non-precipitating clouds the concentration of droplets ranges from $23$ to $1300$ droplets per cubic centimeter~\cite{warner1969microstructure}), the density can be thought of as being approximately constant throughout the cloud the height of the cumulus clouds depends on the amount of moisture in the thermal that forms the cloud, and humid air will generally result in a lower cloud base. In stable air conditions in which their vertical growth is not that high, they are considered to be effectively two-dimensional. \\

In places, cumulus clouds can have holes where there are no water droplets~\cite{warner1969microstructure}. This fact causes to create the fractal structures that it provides powerful tools to classify them in terms of the circumstances in which they form. The fractal structure of clouds has been reported in some previous works \cite{lovejoy1982area,chatterjee1994fractal,madhushani2012fractal}.\\
The self-organized criticality in the atmosphere and clouds was first detected by Peters et al. by analyzing the precipitation \cite{peters2006critical}. They used satellite data and define a critical value of water vapor as a tuning parameter, and precipitation as the order parameter shows a non-equilibrium continuous phase transition to a regime of strong atmospheric convection and precipitation.\\
In an unpolished work, we uncovered the SOC state of clouds directly by analyzing some earth to sky images of cumulus clouds under fair air conditions. The analysis of the level lines of two-dimensional cloud fields strongly suggests that these are in a SOC state, which is also confirmed by a Schramm-Loewner evolution (SLE) analysis, finding that these belong to the sandpile universality class, i.e. $c = -2$ conformal field theory (CFT). Some statistical analysis  has been shown in Fig.(\ref{Fig:cloud})  where $L$ is the box linear size in the box counting method, $l \sim L^{D f^a}$ is the trace length, $R$ is end-to-end distance, $N$ is the number of steps along the trace, $\sqrt{\langle R^2 \rangle} \sim N^{\nu} $, and $ \langle \log l \rangle  = D_f^b \langle \log r \rangle$ in which $r$ is the gyration radius of the loop. Their analysis show $D_f^a = 1.248 \pm 0.006 $, $D_f^b = 1.22 \pm 0.02 $, $\nu = 0.81 \pm 0.01$ where $ \nu = \frac{1}{D_f}$. Also this analysis show that the two-dimensional images of the clouds are very close to the  loop-erased random walkers (LERW) traces with $D_f^a =5/4$.

\subsection{Vibrating Piles}~\label{SEC:Vibrations}

A very important question concerning the SOC model is its stability against external manipulations, like vibrations. As stated above, some experiments have also been done to test the critical properties of real piles in the presence of external vibrations~\cite{jaeger1989relaxation} which were modeled and simulated~\cite{mehta1991vibrated}. Recently we have simulated the BTW model under vibration conditions, affecting the toppling rules in one direction, namely $x$-direction. To this end, we manipulated the toppling rules, which depend on time. The toppling matrix was considered to be
\begin{equation}
\Delta_{(i,j),(i',j')}=\left\lbrace\begin{matrix}
4n &  i=i' ,\ j=j'\\
-n &  i=i',\ j=j'\pm 1\\
-n\left(1+\epsilon_0\sin \omega t\right)  & i=i'+1,\ j=j'\\
-n\left(1-\epsilon_0\sin \omega t\right) & i=i'-1,\ j=j'\\
0 & \text{other}
\end{matrix} \right. 
\end{equation}
where $\omega=2\pi/T$ is the angular frequency, $T$ is the time period, and $\epsilon_0$ is the vibration strength parameter. This toppling rule states that the system is vibrating in the $x$ direction making the model anisotropic. The properties of the model were investigated in terms of $T$ and $\epsilon_0$. We uncovered that the exponents run with $\omega$ and $\epsilon_0$. Importantly increasing the strength of vibrations makes the avalanches more smooth with exponents that depend on the (time and space) directions.

\subsection{Invasion Sandpile Model}\label{SEC:Invasion}

\begin{figure}
	\centerline{\includegraphics[scale=.35]{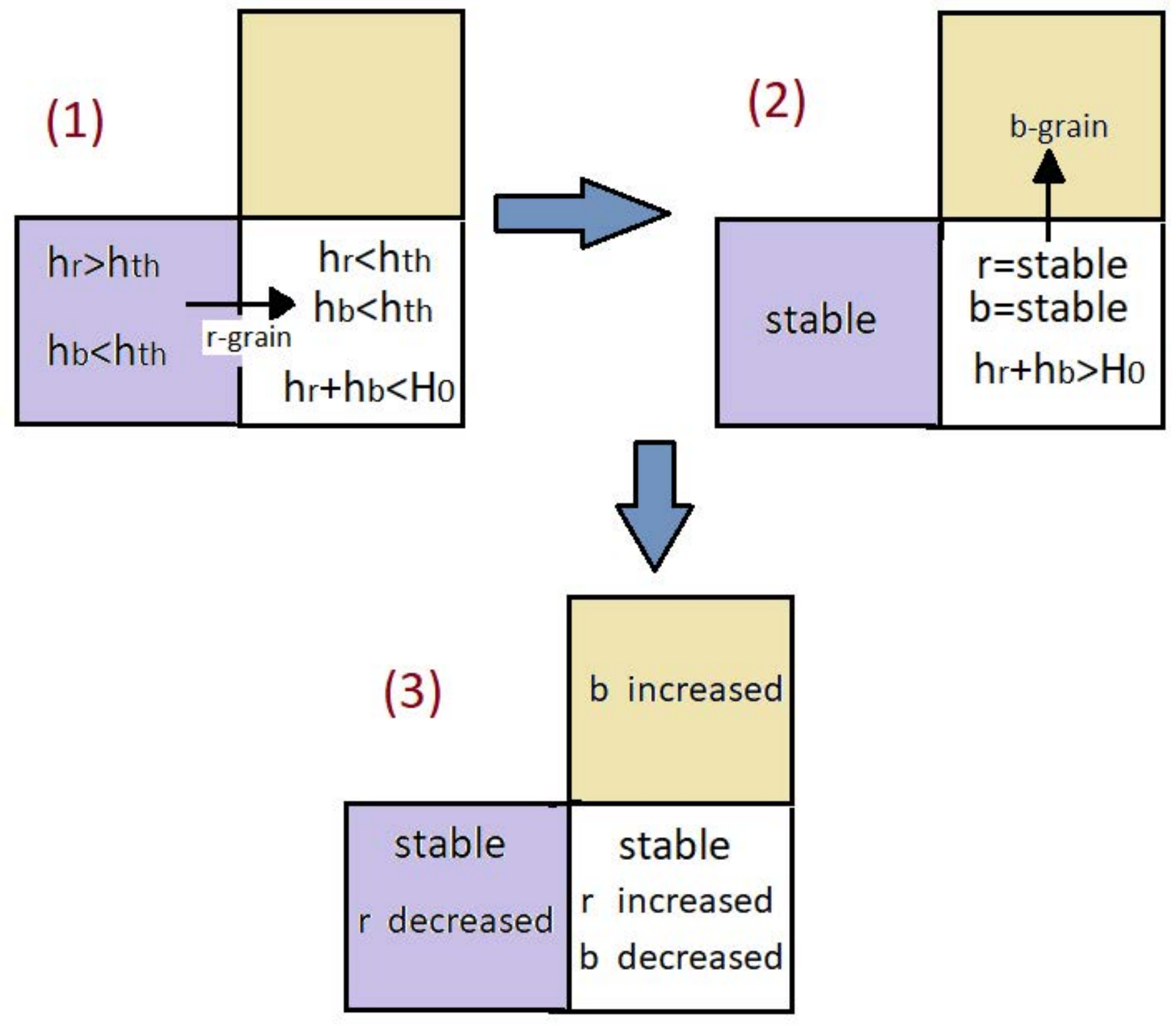}}
	\caption{SEC:~\ref{SEC:Invasion}: Scheme showing the invasion nature of the two-species sandpile model. In $(1)$ the left (blue) site becomes unstable since $h_r>h_{\text{th}}$. In $(2)$ however both grains are lower than $h_{\text{th}}$ for the right (white) site, but $h_r+h_b>H_0$. In this situation $r$ or $b$ is randomly chosen for toppling, here $b$ is chosen. Then in $(3)$ the $b$ content of the upper (dark yellow) site is increased. Therefore, effectively $r$ has invaded $b$ and pushed it towards another site.}
	\label{fig:Schematic}
\end{figure}

In the previous section, we provided evidence that the Darcy model has similarities with the BTW model at the critical occupation. Invasion percolation (IP) is another natural choice, in which one phase invades the other phase towards the production well. In fact, IP~\cite{wilkinson1983invasion} is a standard model to study the dynamics of two immiscible phases (commonly denoted by wet and non-wet phases) in a porous medium~\cite{glass1996simulation,sheppard1999invasion}. During this process, the wet phase invades the non-wet phase, and the front separating the two fluids advances by invading the pore throat at the front with the lowest threshold~\cite{sheppard1999invasion}. This model suffers the lack of the notion of critical saturation introduced in the previous sections. In a parallel line of thinking, one may try to make the BTW model two-phase with the extra tool of invasion.\\
\begin{figure*}
	\centerline{\includegraphics[scale=.13]{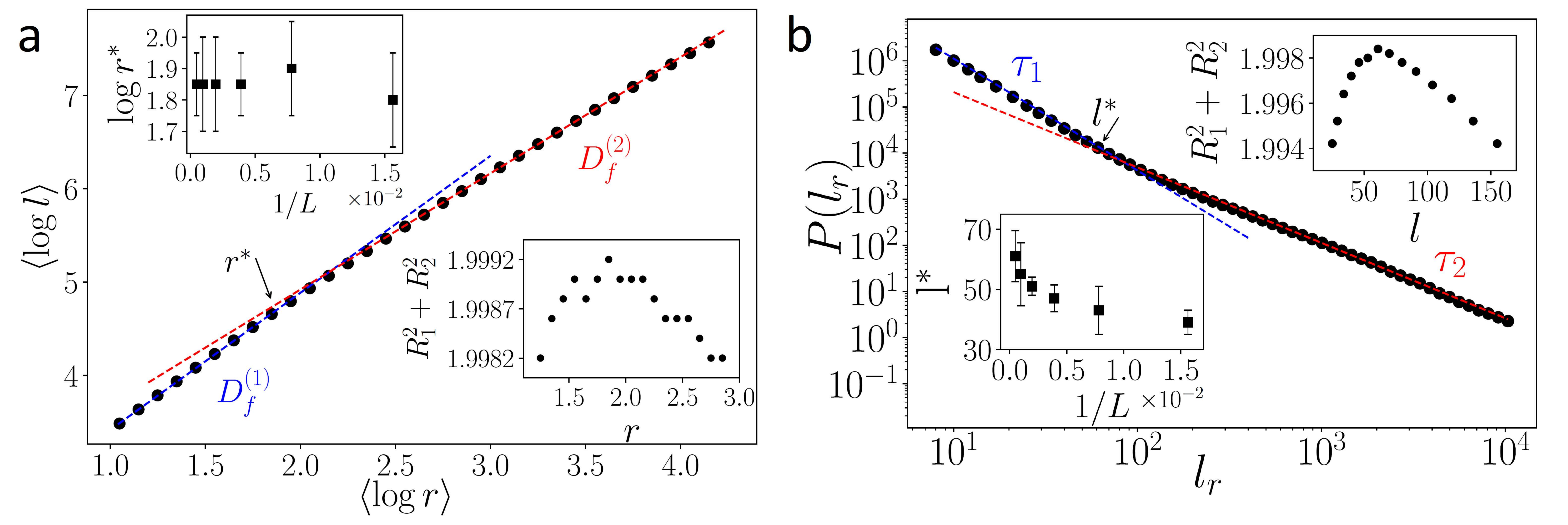}}
	\caption{SEC:~\ref{SEC:Invasion}: (a) $ \langle \log l \rangle $ in terms of $\langle \log r \rangle$ for which the relation gives the fractal dimension $D_f$ . Upper inset shows $r^*$,  the separator of the small and large scale regimes, in term of $L^{-1}$. (b) Distribution function of  $l_r$ (red grains). Bottom inset shows $l^*$ ,the separator of the small and large scale regimes, in term of $L^{-1}$.   Along with the definition of the crossover points by plotting $R_1^2 + R_2^2$
		in terms of the tentative crossover point bottom and upper inset for (a) and (b), respectively.  }
	\label{Fig:IBTW}
\end{figure*}

Let us consider a $L\times L$ square lattice and initially assign two random integers, $h_r$ and $h_b$, to each site, uniformly from the interval $\left\lbrace 1,2,3,..., h_{\text{th}} \right\rbrace $, $h_{\text{th}}$ being the threshold for a one species. $h_r$ and $h_b$ are the number of red and blue sand grains in our two-species sandpile model, representing the two (wet and non-wet) phases in the reservoir. The reported results are independent of the value of $ h_{\text{th}} $, so we set it to $20$. A site $i$ is considered stable if three conditions are fulfilled simultaneously. The first two are the standard ones for the one-species sandpile model, namely, $h_r(i)\leq h_\mathrm{th}$ and $h_b(i)\leq h_\mathrm{th}$, where $h_\mathrm{th}$ represents the CFS. The third one is $h_r(i)+h_b(i)\leq H_0$, where $H_0<2h_\mathrm{th}$ is the second threshold. This additional condition is motivated by the fact that in the non-linear Darcy equations, there is an auxiliary equation expressing that the sum of two-phase saturations $S_w+S_o$ is a constant that depends on the capillary pressure. Thus, a site $i$ is \textit{unstable} and topples if at least one of the following conditions are met:\\
\\
C1: $h_r(i)>h_{\text{th}}$,\\
C2: $h_b(i)>h_{\text{th}}$,\\
C3: $h_r(i)+h_b(i)>H_0$\\
\\
The dynamic goes as follows. Initially, all $ h_r $ and $ h_b $ are chosen randomly from a uniform distribution, such that no site is unstable. Then, iteratively, we first choose a species (either $r$ or $b$, with equal probability) and a site $i$ at random to add a particle of that species, i.e. $h_x(i)\rightarrow h_x(i)+1$ where $x$ is the selected type. If that site becomes unstable, it topples, according to the following rule: If condition C1 is met, then $h_r(i)\rightarrow h_r(i)-1$ and $h_r(j)\rightarrow h_r(j)+1$ where $j$ is the neighbor of $i$ with the lowest red-grain content. If condition C2 is met, then $h_b(i)\rightarrow h_b(i)-1$ and $h_b(j)\rightarrow h_b(j)+1$ where $j$ is the neighbor of $i$ with the lowest blue-grain content. If condition C3 is met, then $h_x(i)\rightarrow h_x(i)-1$ and $h_x(j)\rightarrow h_x(j)+1$ where $j$ is the neighbor of $i$ with the lowest $x$-grain content, and $x$ is randomly chosen to be $r$ (red) or $b$ (blue). As a result of the relaxation of the original sites, the neighboring sites may become unstable and also topple. Therefore, the toppling process is repeated iteratively until all sites are stable again. This collective relaxation is called an \textit{avalanche}. The sand grains can leave the sample from the boundaries, just like in the ordinary BTW model~\cite{bak1987self}. Note that, with two species, an avalanche of one species might trigger an avalanche of the other one, see example in Fig.~\ref{fig:Schematic} and details in the caption. The reason that we call this \textit{invasion} is that here one species pushes the other one due to the finite capacity of the pore, i.e. the total volume of the particles cannot exceeds a threshold (see C3), as in real situations. In the Darcy reservoir model, C3 is an auxiliary equation, where $H_0$ plays the role of the maximum finite saturation that is possible in a pore~\cite{najafi2016water,najafi2014bak} and is the source of the invasion in the invasion percolation model~\cite{wilkinson1983invasion}.\\
\begin{table}
	\caption{SEC:~\ref{SEC:Invasion}: The $\beta$  and $\nu$ exponents from the finite-size analysis, $\tau_1$, and $\tau_2$ for $m$, $s$, $l$, and $r$ corresponding to the red avalanches. The last row contains the exponents for the 2D BTW model for the sake of comparison with $\tau_2(L\rightarrow \infty)$~\cite{Lubeck1997Numerical,najafi2012avalanch}.}
	\begin{tabular}{c | c | c | c | c }
		\hline quantity  & $m$ & $s$ & $l$  & $r$ \\
		\hline $\tau_1(L\rightarrow \infty)$  & $1.04\pm 0.04$ & $0.95\pm 0.03$ & $2.5\pm 0.03$  & $3.1\pm 0.1$ \\
		\hline $\tau_2(L\rightarrow \infty)$ & $1.32\pm 0.02$ & $1.26\pm 0.04$ & $1.63\pm 0.03$  & $1.8\pm 0.1$ \\
		\hline $\beta$   & $-$ & $-$ & $1.87\pm 0.05$  & $1.58\pm 0.03$ \\
		\hline $\nu$     & $-$ & $-$ & $1.21\pm 0.03$  & $0.95\pm 0.03$ \\
		\hline
		\hline $\tau^{\text{2D BTW}}$ & $1.33\pm 0.01$ & $1.29\pm 0.01$ & $1.25\pm 0.03$  & $1.66\pm 0.01$ \\
		\hline
	\end{tabular}
	\label{tab:exponents1}
\end{table}
\begin{table}
	\caption{SEC:~\ref{SEC:Invasion}: The exponents $\beta$, $\nu$, $\tau_1$, and $\tau_2$ for $m$, $s$, $l$, and $r$ corresponding to the two-species avalanches.}
	\begin{tabular}{c | c | c | c | c }
		\hline quantity  & $m$ & $s$ & $l$  & $r$ \\
		\hline $\tau_1(L\rightarrow \infty)$  & $0.95\pm 0.05$ & $0.90\pm 0.05$ & $--$  & $1.61\pm 0.05$ \\
		\hline $\tau_2(L\rightarrow \infty)$ & $1.32\pm 0.02$ & $1.25\pm 0.03$ & $1.50\pm 0.03$  & $2.0\pm 0.1$ \\
		\hline $\beta$   & $-$ & $-$ & $1.90\pm 0.05$  & $1.78\pm 0.03$ \\
		\hline $\nu$     & $-$ & $-$ & $1.18\pm 0.03$  & $0.95\pm 0.03$ \\
		\hline
	\end{tabular}
	\label{tab:exponents2}
\end{table}
 This model has two types of avalanches: one-species and two-species avalanches. The first one involves only the redistribution of grains of one species. In the second, there is mass transport of the two species.  The symmetrical one-species avalanches have the same results for blue and red avalanches. Also, it has two  different regimes: for large avalanche sizes the fractal dimension $D_f^{(2)}$ is consistent with $5/4$ observed for the 2D BTW model \cite{lubeck1997bak}, but for small avalanches, $D_f^{(1)}=1.47 \pm 0.02$ that it is shown in Fig.(\ref{Fig:IBTW}). For extract two regimes the crossover point was used the $R^2$ test \cite{glantz1990primer} ($R^2 =R_1^2+R_2^2 $) that it is shown in bottom inset of Fig.(\ref{Fig:IBTW}a), and the upper inset shows crossover point of fractal dimension as $r^*$ in term of $1/L$ where $L$ is system size. For all measures (gyration radius ($r$), avalanche size ($s$), and avalanche mass ($m$)) is seen this behavior. Table.(\ref{tab:exponents1}) shows all exponent for this measures in one-species avalanches regime. For example in Fig.(\ref{Fig:IBTW}b) is seen  distribution functions of loop length for red grains($l_r$) same Fig.(\ref{Fig:IBTW}a). In the two species avalanches, the regime fractal dimension has two-state. The fractal dimension of large avalanches, $D_f = 1.24 \pm 0.01$, consistent with 2D BTW model. However, the fractal dimension for small avalanches $D_f = 1.31 \pm 0.01$, which is different from the exponent found for one-species avalanches. Table.(\ref{tab:exponents2}) shows all exponent for this measures in two-species avalanches regime. For further information see \cite{najafi2020invasion}.

\subsection{SOC in Exitable Complex Networks}~\label{SEC:ExcitableNodes}

\begin{figure*}
	\centerline{\includegraphics[scale=.15]{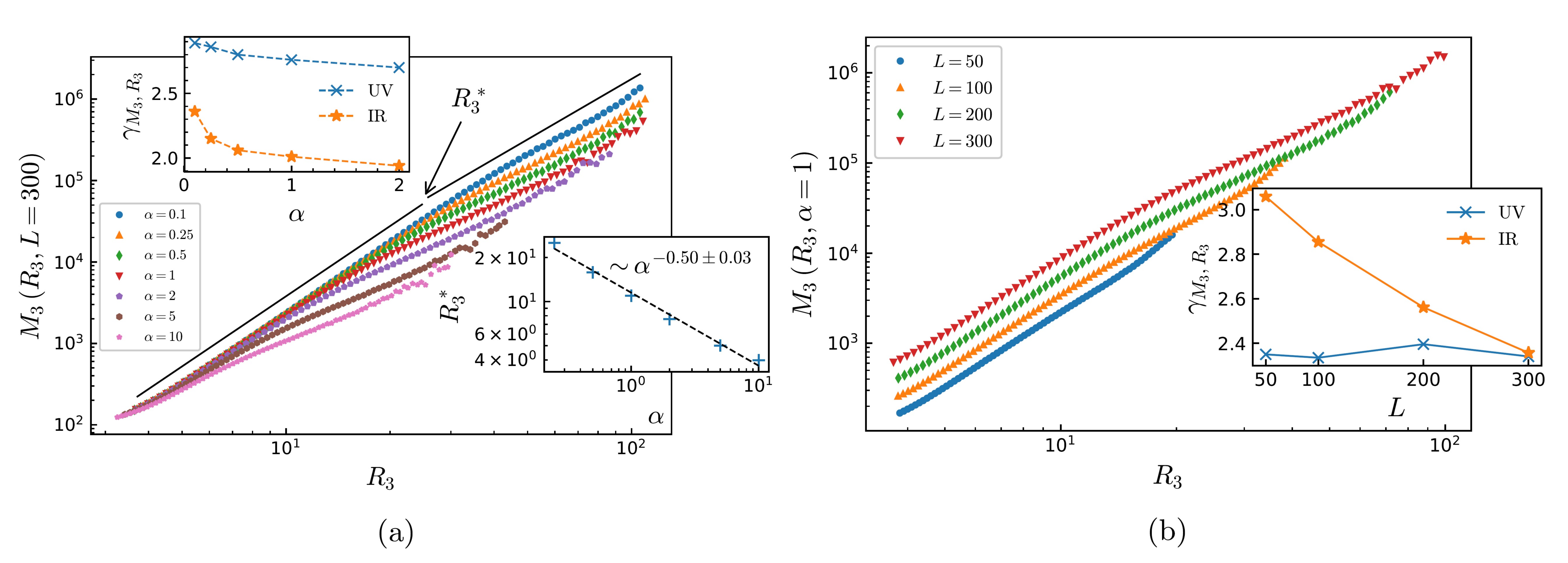}}
	\caption{SEC:~\ref{SEC:ExcitableNodes}: (a) The plot of $M_3$ in terms of $R_3$ and the corresponding exponents $\gamma_{M_3R_3}$. Upper inset: $\gamma_{M_3R_3}^{\text{UV}}$ and $\gamma_{M_3R_3}^{\text{IR}}$. Lower inset: the cross-over radius $R_3^*$ in terms of $\alpha$ with the exponent $0.5\pm 0.03$. (b) The same as (a) for various lattice sizes $L$. The finite size dependent (UV and IR) slopes $\gamma_{M_3R_3}$ have been shown in the inset.}
	\label{Fig:btw_network}
\end{figure*}

\begin{table*}[]
	\begin{tabular}{|c|c|c|c|c|c|c|}
		\hline & $M_3$ & $M_3(0)$ & $R_3$ & $R_3(0)$ & $n_{\text{toppling}}$ & $n_{\text{toppling}}(0)$\\
		\hline $\tau(\alpha=0)$ & $1.34(4)$ & -- & $2.53(5)$ & -- & -- & -- \\
		\hline $\tau_1$ & $1.37(4)$ & $1.6(2)$ & $2.07(5)$ & $2.8(5)$ & $1.33(1)$ & $1.66(2)$ \\ 
		\hline $\tau_2(\alpha=1)$ & $3.46(5)$ & $3.62(9)$ & $5.98(7)$ & $6.1(7)$ & -- & -- \\ 
		\hline $\gamma_{\tau_2}$ & $0.17(3)$ & $0.19(9)$ & $0.18(1)$ & $0.18(3)$ & -- & -- \\ 
		\hline cut$(\alpha=1)$ & $2843$ & $4601$ & $11.2(5)$ & $13.9(4)$ & $3073$ & $6629$ \\
		\hline $\gamma_{\text{cut}}$ & $1.28(9)$ & $1.23(5)$ & $0.42(8)$ & $0.49(7)$ & $1.57(3)$ & $1.06(3)$ \\
		\hline
	\end{tabular}
	\caption{SEC.~\ref{SEC:ExcitableNodes}: The asymptotic values of the exponents. For each quantity there is a "cut" value in which a cross over between small scale behavior (which are consistent with regular 3D BTW model) and large non-universal behavior occur. It has been found these cut-values scale with $\alpha$ in a power-law fashion. For example $M_3^*\equiv M_3^{\text{cut}}=m^{\text{cut}}(\alpha=1)\alpha^{-\gamma_{\text{cut}}}$ which have been shown by "cut" in the table. In contrast to $\tau_1$, $\tau_2$ runs with $\alpha$ for all quantities; $\tau_2=\tau_2(\alpha=1)\alpha^{+\gamma_{\tau_2}}$ which have been shown separately in the table. After~\cite{najafi2018statistical}.} 
	\label{tab:3d-tau}
\end{table*}

Probably the most important reason for considering of the BTW dynamics on top of the complex networks has been the important observation of Beggs et. al.~\cite{beggs2003neuronal} in which it was shown that the propagation of spontaneous activity in cortical networks is self-organized critical phenomena governed by avalanches much similar to the BTW model. The focus of researches in this area had been on the structural and functional properties of random lattice models. One of the most challenges in these systems is finding the circumstances under which the system shows the critical behaviors~\cite{Friedman}. In order to monitor critical behaviors, different time series are usually analyzed in which power-law behavior is expected~\cite{de2006self}. Therefore one may be encouraged to investigate time series of topplings in the SOC model on various random link lattices, like scale free networks~\cite{lee2012sandpiles,lee2004sandpile}, multiplex networks~\cite{lee2012sandpiles}, optimized scale-free network on Euclidean space~\cite{karmakar2005sandpile}, Watts-Strogatz small-worlds~\cite{lahtinen2005sandpiles}, directed small-world networks~\cite{pan2007sandpile}, and on Scale Free Networks with preferential sand distribution~\cite{bhaumik2017sandpile}.\\

An example of the network that is embedded in the Euclidean space is a graph with random links with finite range interaction (RLFRI), i.e. two nodes are connected if their distance is smaller than a control parameter $R$. Therefore the topology of this graph is tuned by $n$ and $R$, where $n$ is the number of links per site. Then dynamic is defined via the following toppling matrix
\begin{equation}
\Delta_{i,j}=\left\lbrace \begin{array}{cc} z_i & \text{if $i=j$} \\ -1 & \text{if $i$ and $j$ are connected} \\  0 & \text{other}\end{array}\right.
\end{equation}
where $z_i$ is the degree of node $i$. \\

The other system of interest is the small world networks that are interpolation between regular and random networks. In addition to the regular links between neighbors, there are some long-range links between random-chosen sites. For this system, we have two dependent further random fields in addition to the height field. The connection matrix $\text{L}(i,j)$ is unity if sites $i$ and $j$ (not neighbors) are connected by a long-range link and zero otherwise. The distribution of lengths and the degree of nodes are chosen to be uniform in the interval of allowed values (naturally the lengths are restricted to the linear size of the system). The other one is $z_c(i)=6+\sum_{j}\text{L}(i,j)$ which accounts for the number of total links in the node $i$.
In this language if the height of a node exceeds $z_i$ it topples according to the rule $h(i)\rightarrow h(i)-\Delta_{i,j}$ in which:
\begin{equation}
\Delta_{i,j}=
\begin{cases} -1 & i \text{ and } j \text{ are neighbors or $\text{L}(i,j)\neq 0$} \\
z_i & i=j \\ 
0 & \text{other.}
\end{cases}
\end{equation}
$\alpha$ (the percent of long range links) is defined as follows:
\begin{equation}
\alpha\equiv 100\times\frac{1/2\sum_{i,j}\text{L}(i,j)}{\text{\# total regular links}}
\end{equation}
In which the factor $\frac{1}{2}$ is to prevent double counting.\\
Fig.~\ref{Fig:btw_network}(a) shows the plot of $\left\langle \log(M_3)\right\rangle$ in terms of $\left\langle \log(R_3)\right\rangle$ whose slope is $\gamma_{M_3R_3}\equiv D_F^{M_3}$ which is the 3D mass fractal dimension for $L=300$ and various $\alpha$'s. We note that $D_F^{M_3}(\alpha=0)\simeq 2.96\pm 0.02$~\cite{dashti2015statistical}. Interestingly it is seen that the graphs smoothly cross over to the large scale regions in which the slope (fractal dimension) ($m_{\text{IR}}\equiv \gamma_{M_3R_3}^{\text{IR}}$) is different from the slope in the small-scale region with the slope $m_{\text{UV}}\equiv \gamma_{M_3R_3}^{\text{UV}}$. We name the small scales as \textit{UV limit} and the large scales as \textit{IR limit}. The point of this behavior change depends on $ \alpha $. This point can easily be calculated using the linear fit of the graphs in each individual region. The transition point ($R_3^*$) is simply the point in which the fits meet each other. The fact that $m_{\text{UV}}(\alpha)$ is nearly $\alpha$-independent and $m_{\text{IR}}(\alpha)$ runs crucially by varying $\alpha$ can be seen in the upper inset of Fig.~\ref{Fig:btw_network}(a). We interpret $R_3^*$ as the point at which the cross-over takes place to the large scale properties since for $r\lesssim R_3^*$ the results are very close to the regular BTW model, whereas for $r\gtrsim R_3^*$ the behavior is different and not universal (presumably mixed with the finite-size effects). More interestingly we have observed that $R_3^*$ is a decreasing function of $\alpha$, i.e. $R_3^*\sim\alpha^{-\zeta}$ in which $\zeta=0.5\pm 0.05$. Since $\alpha$ can be interpreted as the measure of how directly a randomly chosen site is connected to a boundary site at which dissipation occurs, we can say that effectively (on average) a fraction of grains are dissipated in a bulk toppling depending on the amount of $\alpha$. The other effect is the \textit{sink} role of connection sites in micro-avalanches. In fact when $\alpha$ increases, the probability that a micro-avalanche involves a site that has a long-range link to the other micro-avalanches increases. Since roughly speaking, such sites play the role of sink points, one may expect that the effective model for micro-avalanches is a dissipative one. It is known that the dissipative BTW model is equivalent to the massive ghost action
\begin{equation}
S=\int{d^{2}z}\left( \partial\theta\bar{\partial}\bar{\theta}+\frac{m^{2}}{4}\theta\bar{\theta}\right)
\end{equation}
where $\theta$ and $\bar{\theta}$ are complex Grassmann variables and $m^{2}$ is the number of sand grains dissipated in each toppling ($m$ can be fractional). On the other hand it is known that $R_3^*\sim m^{-1}$~\cite{PhysRevE.85.051104}. From these two points, one concludes that effectively our model is equivalent to the dissipative BTW model with $m^2\sim\alpha$. This correspondence is acceptable only for $r\lesssim R_3^*$ and shows that the large scale regime is directly affected by the dissipations in the boundary sites and the finite-size effects. This result is reasonable since the amount of grain dissipation in a single component of an avalanche (the number of sand grains that are transferred out of that area) is proportional to the number of nodes with long-range links in that area. The problem description is not however as simple as stated above since there are surely some other links that return energy to the original micro-avalanche which partly compensates the dissipation effects. It is worth noting a comment concerning the numerical value of $\xi$ which is claimed to be $1/d\approx 0.33$ in three dimensions \cite{bhaumik2013critical} which is true for the total avalanches. For micro-avalanches however, the statistics is different and it is acceptable that $R_3^*$ should be proportional to $m^{-1}\propto\alpha^{-1/2}$ which is representative of the grain dissipation towards the other micro-avalanches and is a well-known property of the dissipative sand-pile models. We have also considered the finite size effect of the results which have been shown in Figs. \ref{Fig:btw_network}(b). The constant trend of $m_{\text{UV}}$ is seen in the inset of this figure, in which it is seen that its numerical value (for $\alpha=1$) is nearly robust against varying lattice size $L$, whereas $m_{\text{IR}}$ changes considerably by lattice size. This reflects the universal behavior of $m_{\text{UV}}$.\\

\subsection{SOC in Imperfect Supports}~\label{SEC:ImperfectSupports}

\begin{figure*}
	\centerline{\includegraphics[scale=.20]{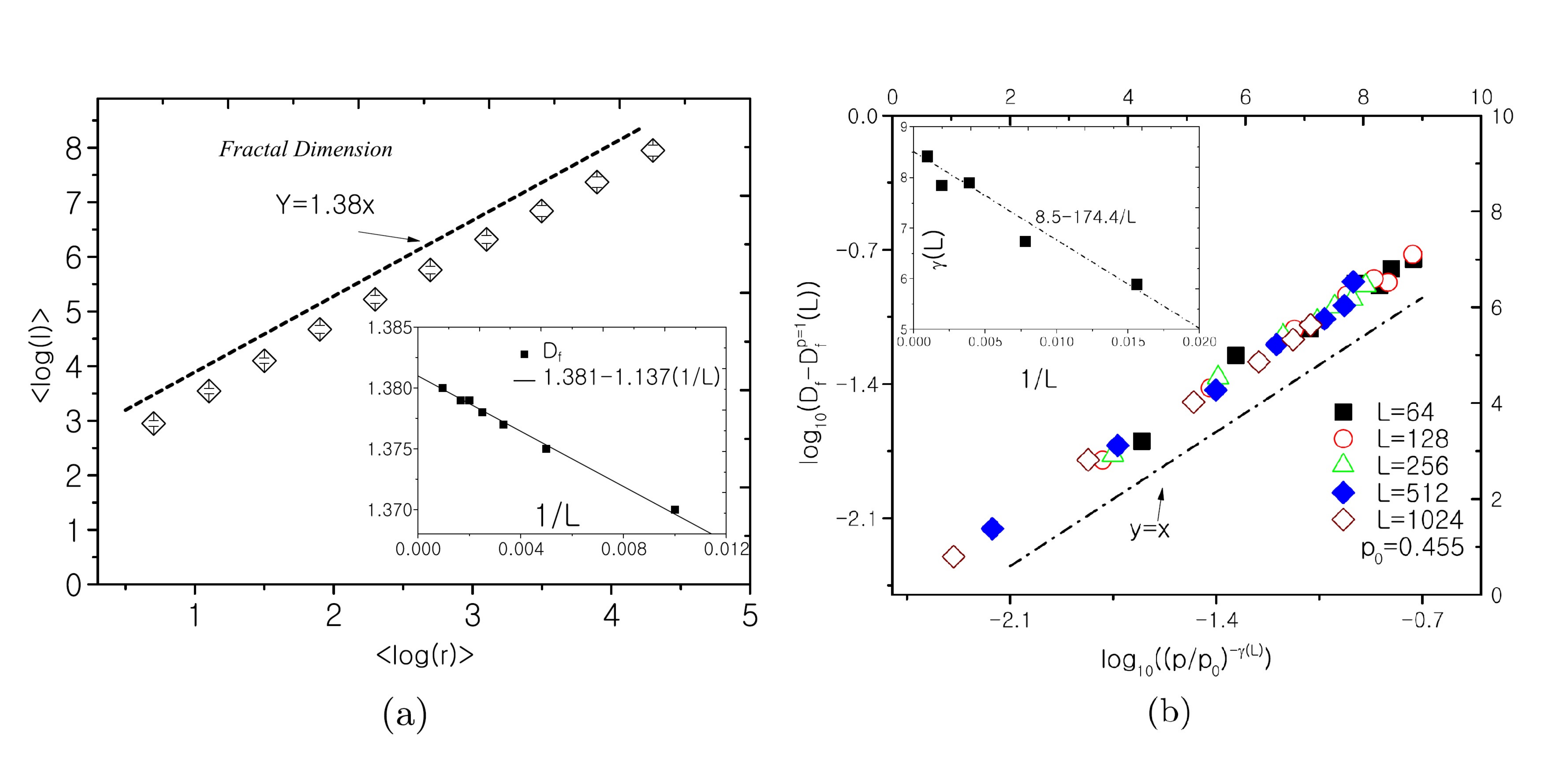}}
	\caption{SEC:~\ref{SEC:ImperfectSupports}: (a) Fractal dimension of the curves for $p=p_c$ and $L = 1024$. Inset shows finite size effect for the fractal dimension, i.e. $D_f^{p=p_c}(L)=1.381 - 1.137(\frac{1}{L})$. (b)  Finite size effect for the fractal dimension for $p>p_c$. Inset shows the dependence of $\gamma( L)$ on $L$}
	\label{Fig:D_f}
\end{figure*}

There are many phenomena in nature in which the dynamics are defined on the percolation lattice which provides motivation for studying this field. Some examples are the recent experiments in which the voids of percolating clusters were filled by (commonly magnetite) nano-particles of ferromagnetic fluids \cite{kose2009label,kikura2004thermal,Mitsuo,kikura2007cluster}, the
loop-erased random walk on the percolation lattices (which is related to watersheds \cite{daryaei2012watersheds}) and fluid propagation in porous media \cite{najafi2015geometrical,oliveira2012post,li2009comparison,wilkinson1983invasion}. Mixing two statistical models (one as the dynamical model and the other as the host for the first one) may be interpreted as the interplay
between two statistical models from which some new non-trivial critical behaviors can emerge. One of the important questions in the contex of self-organized criticality is the effect of stochasticity on its critical behaviors, and on the universality class of the deterministic model. This stochasticity can be annealed (like the Manna model), quenched where the disorder is pinned to lattice points. For the latter case, consider a situation in which some points of the lattice is inaccessible for the sand grains to pass through, namely impermeable sites. One may distribute the impermeable cores throughout the lattice using a single parameter $p$: each site is permeable with the probability $p$, and impermeable with the probability $1-p$. Then the host system is simply described by the percolation theory. In this case, let us define $z_i$ as the number of active neighbors of the site $i$. Then the sand grains are governed by the following toppling matrix: 
\begin{equation}
\Delta_{i,j}=\left\lbrace \begin{matrix}
z_i & i=j\\
-1 & i,j\ \text{are neighbors, and}\ j\ \text{is permeable}\\
0 & \text{other} 
\end{matrix}\right.
\end{equation}
This toppling role has been shown in Fig.(\ref{Fig:BTW}). This model was shown to have the same properties as the Darcy model in 2D (square) lattices at $p=p_c$, whereas they are different for other $p$ values~\cite{najafi2016water,najafi2016bak}. Note that the sand grains are allowed to move on the spanning percolation cluster, that connects two opposite boundaries. In two dimensions, we first construct the percolation lattice as stated in the previous section for various rates of $p$ and then study the BTW model on the percolation clusters.  We start from random $h$ configurations for each amount of $p$ and add sand grains randomly throughout the sample. In this case is seen that after a number of sand grain injections that it depends on the amount of $p$,  the system reaches the steady-state,  e.g. the average height in the percolation lattice ($\bar{h}$ ) becomes constant.\\
Each avalanche has an exterior boundary that forms a loop. To identify these loops, we label each site of the system in each toppling process. it is white if that site is unoccupied or untoppled and black if that site is occupied and toppled. Then, the loops are well-defined as the separators of the black and white sites. For these avalanches, we define gyration radius, loop length, cluster mass, and the number of topplings in each avalanche, represented here by $r, l, m $, and $n_t$ respectively. \\
The samples for $ p \geq p_c$ are self-similar and show critical behaviors, e.g. the fractal dimension is well-defined for them. The dependence of the fractal dimension
on $p$ has been shown in figure \ref{Fig:D_f} in which the finite-size scaling argument has been presented. We also directly observed that the largest system size, the fastest
BTW-like ($p = 1$) behaviors starts. This shows that in the thermodynamic limit the behavior of BTW-like ($p = 1$) is the dominant behavior for all $p_c<p\leq1 $. This hypothesis has also been confirmed by further analysis of e.g. the distribution function of $r, l $ and $m$. We find that the fractal dimension of avalanche exterior boundary (loop) is compatible with the fractal dimension of the exterior boundaries of the geometrical spin clusters of the 2D Ising model,i.e $D_f^{p=p_c}(L = \infty) \simeq D_f^{Ising}= \frac{11}{8}$.\\
\begin{figure*}
	\centerline{\includegraphics[scale=.10]{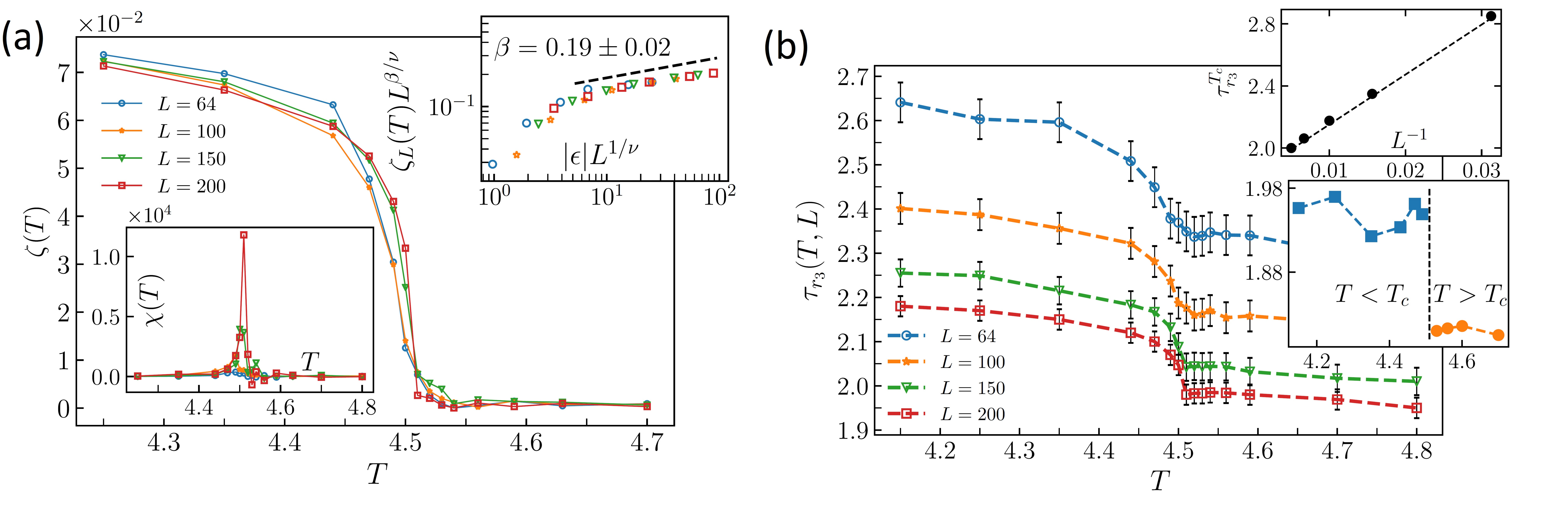}}
	\caption{SEC.~\ref{SEC:ImperfectSupports}: (a) The order parameter $\zeta(T)$ in terms of $T$ for various lattice sizes. Top inset: data collapse of $\zeta$ with exponents $\beta = 0.19 \pm 0.02$ and $\nu=0.75\pm 0.05$ ($\epsilon\equiv \frac{T-T_c}{T_c}$). Lower-right inset: $\chi(T)\equiv\partial \zeta/\partial T $ in terms of $T$, showing a peak at $T_c$. Lower-left inset: average coordination number $\bar{Z}$ of the Ising clusters as a function of $T$ for various systems sizes. (b) The critical exponent of the distribution function of gyration radius $\tau_{r_3}$ in terms of $T$. Upper inset: the finite size dependence (linear in terms of $L^{-1}$) of $\tau_{r_3}$ at $T_c$, revealing that $\tau^{T_c}_{r_3}(L\rightarrow\infty)=1.86\pm 0.03$. Lower inset: the same finite size extrapolation of $\tau_{r_3}$ for all temperatures. The exponent undergoes a clear jump at $T_c$.}
	\label{Fig:3dising}
\end{figure*}
In this part along with the three-dimensional analysis, we study the energy propagation through two-dimensional slices, i.e. cross-sections of the three-dimensional system. The problem of two-dimensional propagation of sand
grains (energy) in three dimensional systems seems to be very important from both theoretical and experimental sides. More precisely the important question in the theoretical physics is how the information in $d + 1 $dimensions would be reflected in its $d$ dimensional sub-system. For this purpose one should map the original $d + 1$ dimensional model to a d-dimensional one and measure how some information is lost and how the degrees of freedom in the subtracted dimension affect the $d-$dimensional model, i.e. which model lives in the lower dimensional system.
We had two separate studies: critical $p = p_c$ and off-critical $p_c < p ≤ 1 $regimes and the fractal dimensions and the distribution functions of various statistical observables have been studied vie the moment analysis. For the critical case, some proper finite-size scaling relations were observed and some resulting exponents were observed to be compatible with the 2D BTW model. The exponents of the quantities in 2D cross-sections are compatible with the 2D Ising universality class. These exponents satisfy also some hyper-scaling relations. For the off-critical case in three dimensions we have observed that the exponents change logarithmically with $p − p_c$ violating the hyper- scaling relations obtained for the critical case. For the 2D induced model in the off-critical regime, we showed that there is a p-value ($p_0 \in(0.5,0.6))$ at which the behavior of the system changes. This is reminiscent of the previously observed occupation number at which the percolation probability becomes maximum in the BTW model on the 2D site-diluted percolation lattice \cite{najafi2016bak}. We conclude that the system for $p_c^{3D} \leq p < p_c^{2D}$ in the cross-sections does not have a thermodynamic limit, whereas for $p \geq p_c^{2D} $the system is identical to the $p = 1$ system. For details see \cite{dashti2015statistical,najafi2018sandpile}
\\
The above analysis was for the case where the imperfections are uncorrelated. Turning on the correlations for the spatial configuration of the imperfections over the lattice makes the results different. \\
Based on the determination of the fractal dimension of the external perimeter of the avalanches, and the Schramm-Loewner evolution, it was suggested in~\cite{najafi2018coupling} that the BTW on the critical percolation, results to critical Ising universality class, and also the BTW model on the Ising-correlated percolation lattice results to self-avoiding walk universality class. \\
More interesting is the BTW model on the three-dimensional Ising correlated lattice, for which the magnetic phase transition is not accompanied by a percolation transition, allowing us to measure the properties of the model across the transition point. In this case, we implement the dynamics of the Bak-Tang-Wiesenfeld model (BTW)~\cite{Bak1987Sel}, also known as the Abelian sandpile model, on a diluted cubic lattice. This lattice is comprised of sites that are either active (through which sand grains can pass) or inactive (completely impermeable to sand grains), which are labeled by the quenched variable $s$ (called \textit{spin}) that is $+1$ for the active case, and $-1$ for the inactive one. We use the Ising model at finite temperatures ($T$) to obtain the \textit{spin} configuration, which is expressed by the Hamiltonian $H=-J\sum_{<i,j>}s_is_j$, where $s_i$ is the spin on site $i$, $J>0$ the ferromagnetic coupling constant, and $\left\langle i,j\right\rangle $ means that $i$ and $j$ are nearest neighbors. The 3D Ising model undergoes a magnetic phase transition at $T=T_c\approx 4.51$ for the cubic lattice. Since the spin clusters (two sites belong to the same cluster if they are nearest-neighbors and have the same spin) of the 3D Ising model on the cubic lattice \textit{percolate at any temperature}, no percolation transition takes place at $T_c$. This can be understood by noting that the critical site percolation threshold for the cubic lattice is around $0.32<0.5$($=$ occupancy probability for $T\rightarrow\infty$ of the Ising model). After constructing an Ising configuration at a given temperature using Monte Carlo, we implement the BTW dynamics on top of the spanning (majority) spin cluster (SSC), i.e. a cluster comprised of spins with the same orientation connecting two opposite boundaries of the lattice. Free boundary conditions are imposed in all directions. In the BTW dynamics, we consider on each site $i$ a height $h_i$ (the number of sand grains) taking initially randomly (independently and uncorrelated) with the same probability one integer from $\left\lbrace 1,...,Z_i\right\rbrace $, in which $Z_i$ is the number of active neighbors of the $i$th site. Then we add a sand grain at a random site $i$, so that $h_i\rightarrow h_i+1$. If this site becomes unstable ($h_i>Z_i$), then a toppling process starts, during which $h_j\rightarrow h_j-\Delta_{i,j}$, where $\Delta_{i,j}=-1$ if $i$ and $j$ are neighbors, $\Delta_{i,j}=Z_i$ if $i=j$, and is zero otherwise. After a site topples, it may cause some neighbors to become unstable and topple, and so on, continuing until no site is unstable anymore. Then another random site is chosen and so on. The average height grows with time until it reaches a stationary state after which the number of grains that leave the system through the boundary is statistically equal to the number of added ones. The dynamics can be implemented with either sequential or parallel updating. Criticality in three dimensions also induces  two-dimensional (2D) critical properties, which enables us to apply 2D techniques like conformal loop ensemble theory~\cite{dashti2015statistical,najafi2018statistical,najafi2018sandpile,dashti2017bak}. Here we consider three-dimensional (3D) avalanches, as well as their two-dimensional (2D) projections on the horizontal plane.\\

To characterize more precisely the two SOC phases, we analyze the average number of topplings per site (toppling density) in avalanches. We may define as the order parameter $\zeta(T) \equiv f_{\text{perc}}-f(T)$, where $f(T)=\frac{m(T)}{N(T)}$, $m(T)$ being the number of topplings, $N(T)$  the number of sites in the spanning (majority) spin cluster, and $f_{\text{perc}}\equiv f(T=\infty)$. 
Fig.~\ref{Fig:3dising}(a) reveals that $\zeta(T) $ is zero for $T>T_c$, and starts to grow continuously in a power-law fashion when $ T$ is decreased below $T_c$  signaling a phase transition at $T=T_c$, at which $\chi(T)\equiv \frac{\partial \zeta}{\partial T}$ shows a distinct peak. The finite size scaling relation for $\zeta$ is $\zeta_L(T)=L^{-\beta/\nu}G_{\zeta}\left(\epsilon L^{1/\nu} \right)$ (see upper inset of Fig.~\ref{Fig:3dising}(a)) in which $\epsilon\equiv \frac{T-T_c}{T_c}$, $G_{\zeta}(x)$ is a scaling function with $G_{\zeta}(x)|_{x\rightarrow\infty}\rightarrow x^{\beta}$, and $\beta=0.19\pm 0.02$ and $\nu=0.75\pm 0.05$ are the resulting critical exponents. The case $T\rightarrow\infty$ corresponds to a site percolation cluster with occupation probability $p=\frac{1}{2}$. For $T>T_c$, therefore we will call this phase SOC$_{p=\frac{1}{2}}$. The behavior in the $T<T_c$ region, however, is dominated by $T=0$ i.e. the regular lattice, and therefore we call this phase BTW.
At the transition point all of these exponents show a sharp change. For example, $\tau_{r_3}$ in the lower inset of Fig.~\ref{Fig:3dising}(b) abruptly changes its value from BTW to SOC$_{\frac{1}{2}}$ at $T=T_c$, its value being completely different for $T<T_c$ and $T>T_c$. It is step-like in the limit $L\rightarrow \infty$ which is obtained by linear extrapolation in terms of $1/L$. We observed that for all temperatures $T<T_c$, $\tau_{r_3}$ extrapolates to $1.94\pm 0.04$, and for $T> T_c$, it is $1.76\pm 0.04$. At $T=T_c$, this exponent is $1.86\pm 0.03$ which is different from both values.

\subsection{Diffusive Sandpiles}~\label{SEC:Diffusive}

\begin{figure*}
	\centerline{\includegraphics[scale=.15]{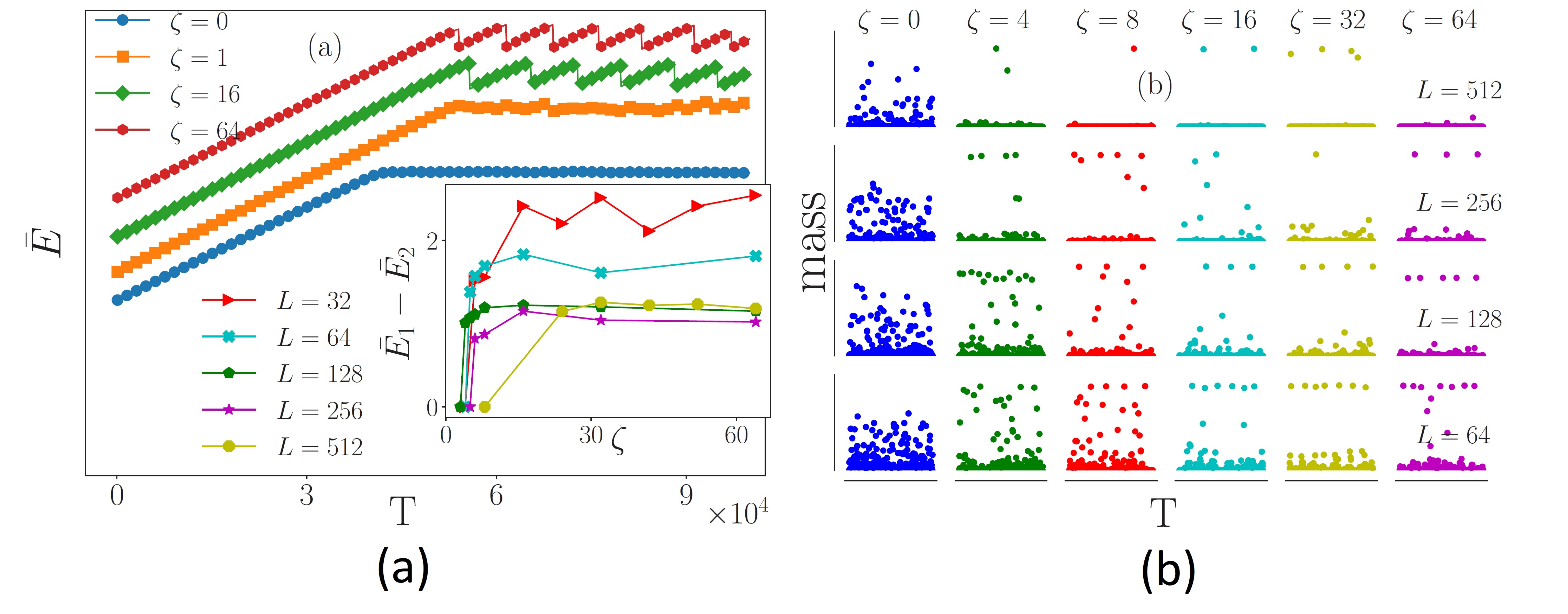}}
	\caption{SEC.~\ref{SEC:Diffusive}: (a) The (shifted) height average in terms of $T$ (the number of injections) for various rates of $\zeta$. Inset: $ \bar{E}_1 − \bar{E}_2  $in terms of $\zeta$. (b) Mass
		as a function of time $T$ for various rates of $\zeta$ and $L$.}
	\label{Fig:dbtw}
\end{figure*}

let us introduce the local smoothings. A local smoothing is defined as the action in which a site is chosen randomly and is checked for a more stable configuration. To do this, the height of its neighbors is checked. Suppose that the selected site is $i$ with nearest neighbors $i_1$, $i_2$, $i_3$ and $i_4$. Among these neighbors, label $i_{\text{max}}$ as the site in which $E(i_{\text{max}})=\text{Max} \left\lbrace E_i \right\rbrace_{i=1}^{4} $, and $i_{\text{min}}$ as the site in which $E(i_{\text{min}})=\text{Min} \left\lbrace E_i \right\rbrace_{i=1}^{4}$. A local smoothing is composed of two updates: $\delta E_1\equiv \text{int}\left[(E(i_{\text{max}})-E(i))/2 \right]$ grains (if positive) flow from the site $i_{\text{max}}$ to $i$, and then $\delta E_2\equiv \text{int}\left[(E(i)-E(i_{\text{min}}))/2 \right]$ grains (if positive) flow from the site $i$ to $i_{\text{min}}$, in which $\text{int}[x]$ is the integer part of $x$. In case of more-than-one sites having the same (maximum or minimum) height, the site from/into which the grains flow is chosen randomly. If the site $i$ is locally maximum (minimum), automatically no grain flows into (from) the site to the neighbors. No local smoothing is applied to the unstable sites. Therefore, we have two kinds of relaxations: the sites which are unstable topple and the sites which are chosen for local smoothings moderate their local height gradient. We call the first procedure as the \textit{toppling} and the second one as the \textit{local smoothing}. This problem can also be called a \textit{diffusive sandpile model}, in which the grains are lubricated such that they have the chance to slip to the neighboring sites. more detail  in  ~\cite{PhysRevE.99.042120}\\

Let us consider the problem in the mean-field (MF) level. Consider a square lattice with $N=L^2$ sites and $4L$ boundary sites. Suppose that the avalanche mass (number of toppled sites in an avalanche) at time $T$ had been $A(T)$, and one energy unit is added at $T+1$, and also the average height at time $T$ ($T$th injection) is considered to be $\bar{E}(T)$. When one energy unit is added to the system, then the average energy is raised by $1/N$. But some energy is dissipated from the boundaries. In the mean-field level, we should first calculate the probability that a boundary site had been unstable in the previous avalanche. This probability is simply the number of boundary sites ($4L$) times the probability that a randomly chosen site is involved in the avalanche. The latter is equal to $A(T)/N$. All in all, we reach the following result for the time-dependent average energy:
\begin{equation}
\bar{E}(T+1)=\bar{E}(T)+\frac{1}{N}-4L\frac{A(T)}{N}.
\end{equation}
The above analysis reveals that in the steady-state in which $\bar{E}(T+1)=\bar{E}(T)$, we have $\bar{A}(T)=\frac{1}{4L}$.  Now consider the effect of local smoothing. We suppose that its effect is decreasing the range of the corresponding avalanches $A'(T)$, and also suppose that $A'(T)$ is proportional to $A(T)$ (the same avalanche in the absence of local smoothing). The proportionality constant is surely $\zeta$-dependent, i.e. $A'(T)=f(\zeta)A(T)$. In this case, under the conditions that the zero-smoothing $\zeta=0$ system is in the steady-state, we have:
\begin{equation}
\bar{E}(T+1)=\bar{E}(T)+\frac{1}{N}(1-f(\zeta)).
\end{equation}
This means that $\bar{E}$ grows linearly with time (with the proportionality constant $\frac{1}{N}(1-f(\zeta))$. This growth takes place up to a time that $\bar{E}\rightarrow E_{th}$ at which the avalanche mass reaches the system size. In this case, $A(T)\approx N=L^2$. In this case the first equation of this document shows that 
\begin{equation}
\bar{E}(T+1)=\bar{E}(T)-(4L-1).
\end{equation}
In this case, the average energy decreases abruptly by $4L-1$. To continue, we introduce the probability of $\bar{E}(T+1)=z$, conditioned to $\bar{E}(T)=M$, which is found to be:
\begin{equation}
P(\bar{E}_{T+1}=z|\bar{E}_T=M) = \left\lbrace \begin{array}{ll}
\delta_{z,M+\frac{1}{N}(1-f(\zeta))} & M<E_{th}\\
\delta_{z,M-4L+1} & M\approx E_{th}
\end{array}\right. 
\end{equation}
One of the important quantities in the analysis of the dynamical systems is the branching ratio defined by $b(M)\equiv\textbf{E}\left[\frac{\bar{E}_{T+1}}{M}|\bar{E}_T=M \right]$. For a given $M$, if $b(M)>1$ then the average $\bar{E}$ grows, and if $b(M)<1$ then it decreases. The above calculations show that 
\begin{equation}
b(M)=\left\lbrace \begin{array}{ll}
1+\frac{1-f(\zeta)}{NM} & M<E_{th} \\ 1-\frac{4L-1}{M} & M\approx E_{th}
\end{array}\right. 
\label{Eq:meanfield}
\end{equation}
in which $\textbf{E}\left[ \ \right]$ is the expectation value. Note that when $b(M)>1$ ($b(M)<1$), then for a given $M$, the average number of grains of the system will increase (here linearly) (decrease, here abruptly) with $T$. This relation predicts that a bifurcation takes place at a non-zero $\zeta$, above which some oscillations occur. For the first branch, the mean height increases linearly with $T$ up to the time at which $\bar{E}\approx E_{th}$. At this point, the average height drops abruptly by $\delta \bar{E}\approx -4L$ (the lower branch). This is accompanied with some large avalanches, which are named as \textit{spanning avalanches (SA)}. This dropping should be independent of $\zeta$.
To test these predictions, we have calculated and plotted $\bar{E}$ in terms of $T$ for various rates of $\zeta $in Fig. \ref{Fig:dbtw}(a). Two separate regimes are distinguishable in this figure: In the primitive times it increases linearly, and for large enough times it enters a new regime, e.g., for $\zeta = 0$ it is nearly constant. However, for nonzero $\zeta$ we see that some oscillations arise in which the average grain number drops abruptly after a linear part in accordance with the prediction of the MF approach. In the inset of this figure, we have plotted the difference between these two limits (among which $\bar{E}$ oscillates)
$\bar{E}_1 - \bar{E}_2$ in terms of $\zeta$, which quantifies these oscillations. This figure characterizes the bifurcation point at which the transition to the oscillatory regime takes place. Actually $\bar{E}_1 - \bar{E}_2$ starts from zero in small enough $\zeta$ and at some $L$-dependent bifurcation point ($\zeta^* $) grows rapidly, and then saturates immediately.\\
Fig. \ref{Fig:dbtw}(b) visualizes this event, in which a new characteristic reference point appears for large enough $\zeta$. In this figure, we have shown the mass of the avalanches (≡ the number of distinct toppled sites in the avalanche) as a function of time for various rates of $\zeta$ and $L$. Consider for example $L = 64$ in the regime $ ζ \gtrsim 8$, for which the masses of some avalanches reach the system size, i.e., the top points in the figure whose mass is almost $(64)^2 = 4096$. These avalanches are the mentioned SAs and are absent in small $\zeta$. The SAs and the abrupt drop of average height occur simultaneously and therefore have the same origin (both belong to the lower branch of Eq.(\ref{Eq:meanfield})). The avalanches that belong to the first branch, whose mean sizes grow linearly with the injections are called deformed avalanches (DAs). The mean size of DAs depends on $E_{th} −\bar{E}$. It is notable that the microstates which grow with time in
the observed quasistationary state are transient. The existence of transient states in the quasi steady state may lead to new studies, and new insights may come up for the phenomena of SOC as a whole.

\subsection{propagation of electrons in 2D electron gas}~\label{SEC:2DEG}

$1/f$ is a well-observed phenomenon in condensed matter systems, especially in two-dimensional electron gas (2DEG). Whether its origin is in avalanche-like dynamics or not need a detailed analysis of the system. Here we report on this possibility, by introducing an avalanche base model for two-dimensional electron gas. \\
In~\cite{najafi2018percolation} the author presents a possibility based on which the percolation of the electrons with avalanche dynamics can be a source for the MIT of the two-dimensional electron gas in zero magnetic fields. They call it the \textit{semi-classical localization} of electrons, which corresponds to \textit{percolative}-\textit{non-percolative} phase transition, although it is different from the conventional percolation theory in essence. Although we don't have enough reasons to call it SOC, since the dynamics are much similar to BTW-like avalanches, we brought it in this section. The \textit{percolative} phase has the property $\frac{\text{d}}{\text{d}T}\sigma<0$ ($\sigma\equiv$ the conductivity) which is the characteristics of the metallic phase. Interestingly this MIT occurs in the diffusion regime of 2DEG and therefore has nothing to do with the Anderson localization. In this model, we consider a two-dimensional electron gas in contact with some electronic reservoirs. The dynamics of the electrons are divided in two categories according to the phase relaxation time $\tau_{\phi}$ associated with inelastic or spin-flip scattering up to which the electrons retain their coherence. Corresponding to this, we divide the spatial dynamics of the electrons to two scales: $l \ll r \ll \l_{\phi}$ and $r\gg l_{\phi}$ in which $l$ is the mean free path due to the electron-electron or the electron-phonon interactions, $l_{\phi}\equiv \sqrt{D\tau_{\phi}}$ is the phase relaxation length, $r$ is the length scale of the electron dynamics in time $t$ which can be estimated classically as $r\sim\sqrt{Dt}$ and $D$ is the diffusion coefficient. In the first scale the electrons retain their quantum phase, whereas, for the latter case, the picture can be semi-classical, since quantum fluctuations in this scale do not play a vital role and one can use the classical Boltzmann transport equation~\cite{altshuler1985electron}. This approach has been proved to be useful in many situations and physical interpretation of some phenomena, like the interpretation of finite-size power-law conductivity of 2DEG~\cite{backes2015observation}, the self-averaging~\cite{bruus2004many}, and the percolation prescription of 2DEG~\cite{meir1999percolation} each of which considers the linear size $\Delta L\sim\l_{\phi}$ as an important spatial scale. We have treated the electron gas inside these cells purely quantum mechanically, but for the transport of the particles to the neighboring cells some semi-classical rules have been developed. To be most symmetric, the cells have been chosen to be hexagonal. This model is essentially different from conventional percolation theory (used for example in Ref.~\cite{meir1999percolation} and Ref.~\cite{sarma2005two}). It is most suitable to be called a \textit{cellular automaton model} in which some electrons can propagate throughout the system according to some local dynamical rules. In this model, the electrons propagate through the system according to the (temperature-dependent) energy content and also the chemical potentials of the cells. In some cases, some electrons can reach from one side to the opposite boundary, which is called \textit{percolated}.\\
\begin{figure*}
	\begin{subfigure}{0.33\textwidth}\includegraphics[width=\textwidth]{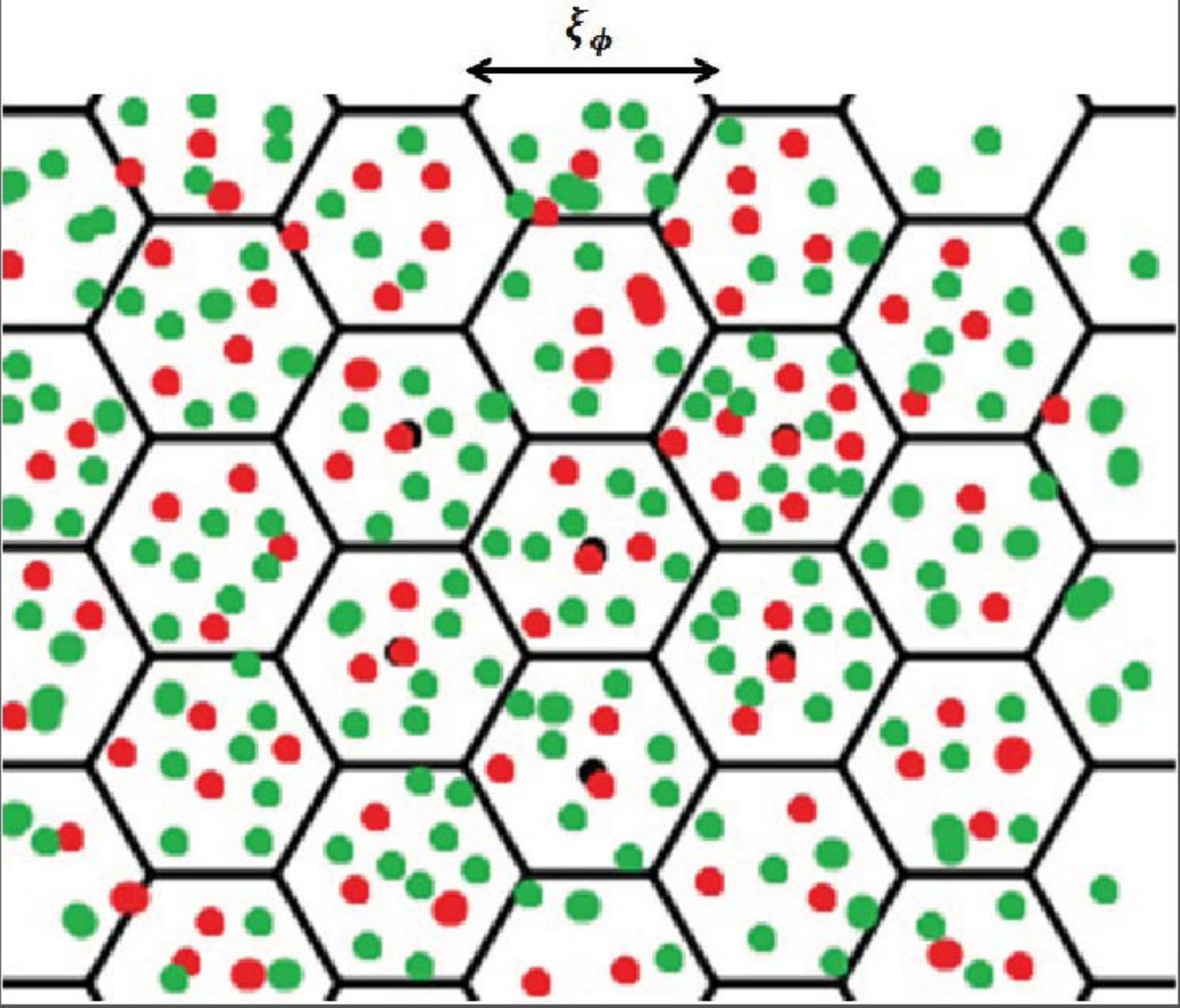}
		\caption{}
		\label{Honeycomba}
	\end{subfigure}
	\begin{subfigure}{0.33\textwidth}\includegraphics[width=\textwidth]{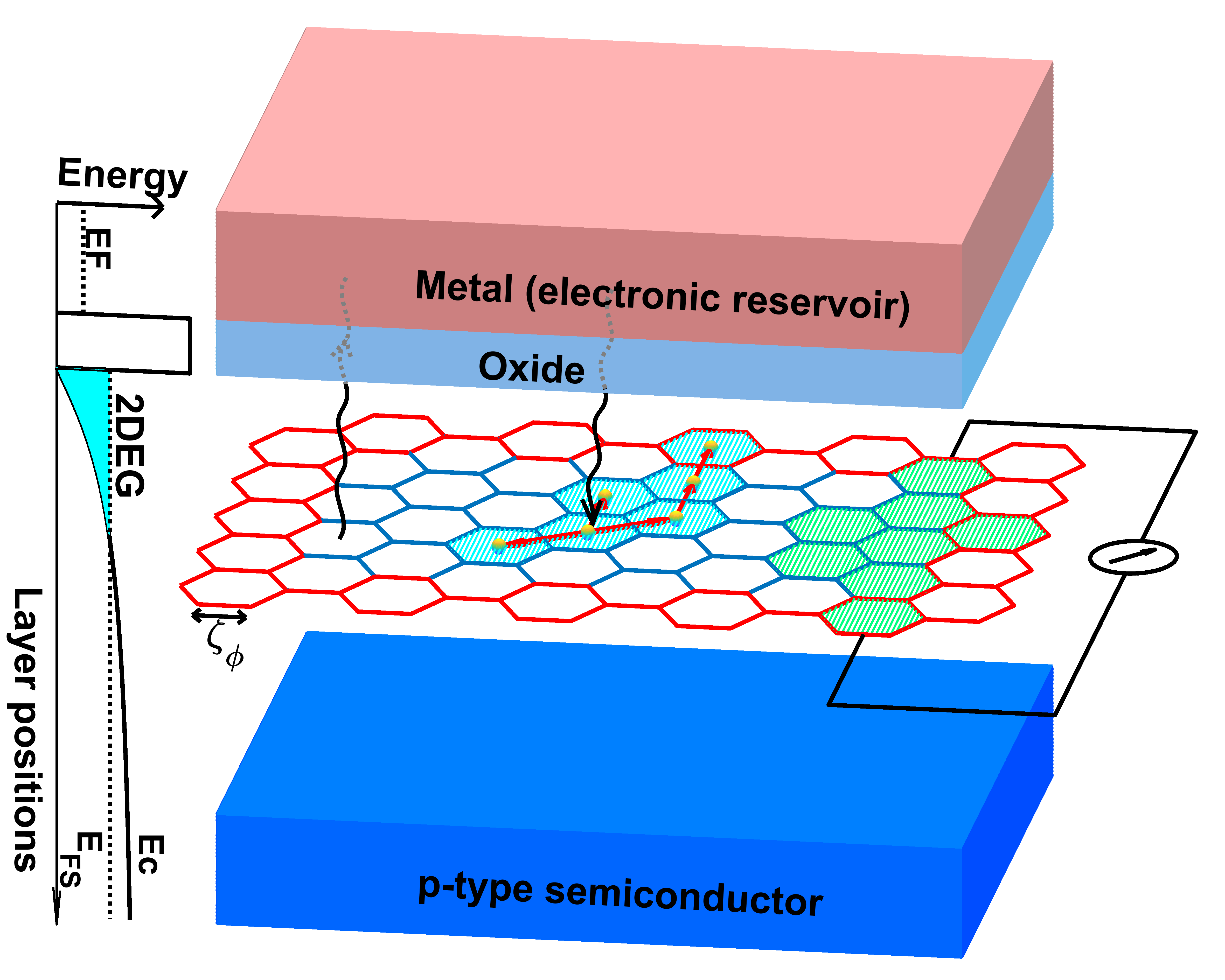}
		\caption{}
		\label{Honeycombb}
	\end{subfigure}
	\caption{SEC.~\ref{SEC:2DEG}: (a) A schematic graph of dividing the 2D lattice into many hexagons. Inside the hexagons we have a pure quantum electron gas. The transfer between the cells occurs semi-classically. The green points show electrons and the red ones show the impurities. (b) A schematic set up of a 2D electron gas surrounded by charge reservoirs. The same partitioning has been carried out in this case. The electron can enter and exit the 2D system at any random point (with some energy considerations), e.g. from the boundaries. }
	\label{universal-resist}
\end{figure*}
\begin{figure*}
	\begin{subfigure}{0.27\textwidth}\includegraphics[width=\textwidth]{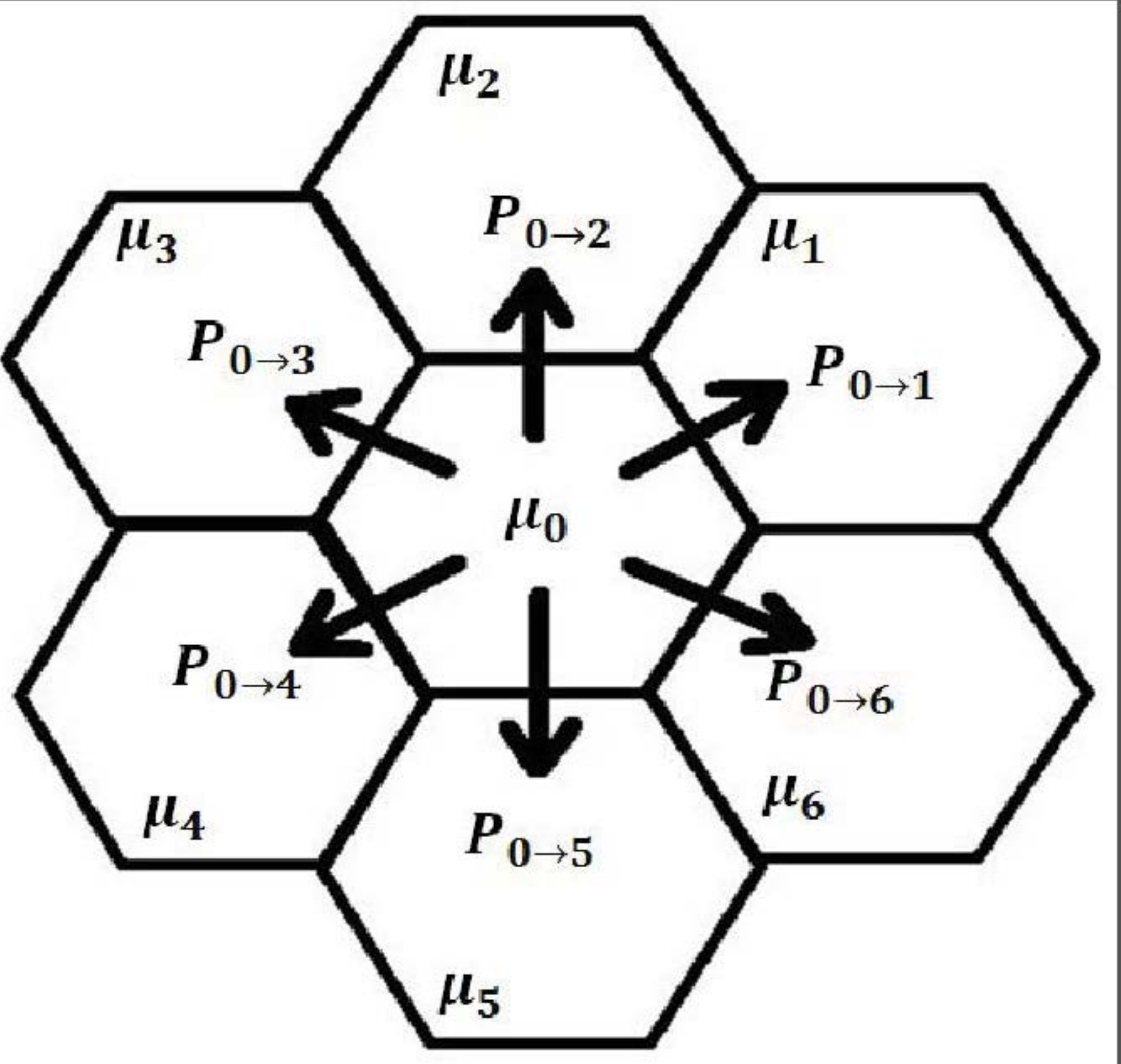}
		\caption{}
		\label{Fig:Transport}
	\end{subfigure}
	\begin{subfigure}{0.40\textwidth}\includegraphics[width=\textwidth]{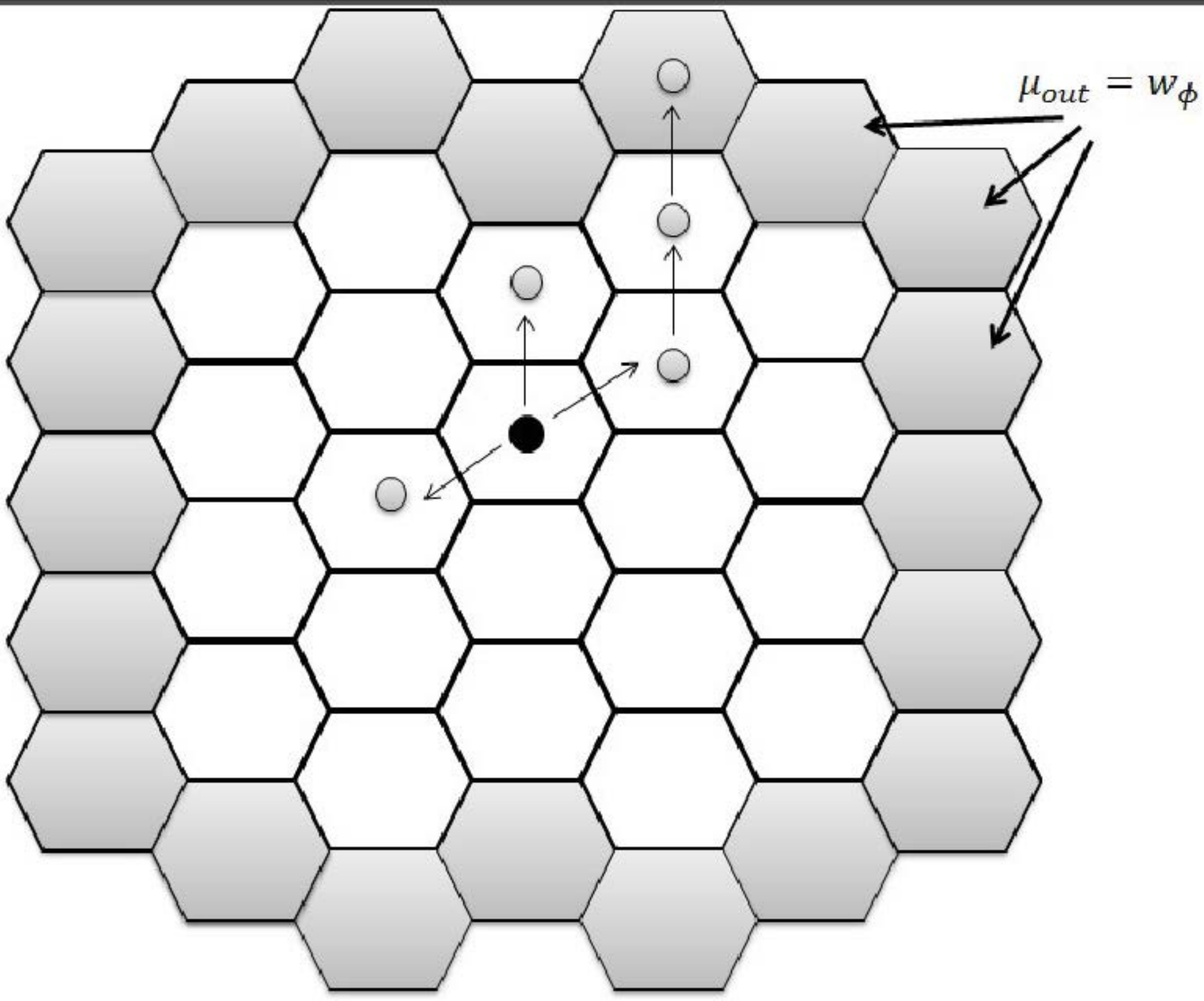}
		\caption{}
		\label{Fig:Movement}
	\end{subfigure}
	\caption{SEC.~\ref{SEC:2DEG}: (a) The electron transition to the neighboring cells. (b) An schematic movement pattern of an electron in the virtual lattice. The gray cells are boundary sites (outer sites with $\mu=w_{\phi}$) from which the electrons can leave the system. The black circle is the site at which the electron has been injected and the gray circles represent the sites which have been unstable and relaxed through a chain of charge transfers}
	\label{MovementPattern}
\end{figure*}

\subsubsection{general set up}

The energy of the electron gas and the chemical potential inside each cell is calculated by means of the Thomas-Fermi-Dirac (TFD) approach. The average energy of the $i$th cell, inside which the charge is supposed to be uniform, is $\left\langle E_i \right\rangle=K(T,\tilde{N}_i)+V_{ee}(T,\tilde{N}_i)+E_{\text{imp}}(T,\tilde{N}_i)$ in which the terms are finite temperature averages of the kinetic, the electron-electron interaction and the impurity energies respectively and $\tilde{N}_i$ is the number of electrons in the cell. The total energy of a cell is shown to be~\cite{najafi2018percolation}:
\begin{equation}
E_T=\sum_{i=1}^{L}\left[-\alpha T^2\text{Li}_2\left(1-e^{N_i/T}\right) +\beta_0 N_i^2-\gamma_i N_i\right]
\end{equation}
in which $\alpha=2m\left( \frac{\pi k_B \zeta_{\phi}(T)}{\hbar}\right)^2$, $\beta_0=\frac{1}{8\sqrt{2}\epsilon_0 \zeta_{\phi}(T)}\left( \frac{e\alpha}{k_B}\right)^2$, $\gamma_i=\text{sinh}^{-1}(1)\frac{\alpha e^2}{\pi\epsilon_0k_B \zeta_{\phi}(T)}Z_i$, $N_i=\frac{k_B}{\alpha}\tilde{N}_i$ and $L$ is the total number of cells. Apparently this calculation contains some simplification which takes a part some complexities that are unnecessary for the physics of the proposed MIT. \\
The chemical potential of a cell ($\mu_i=\partial A_i/\partial N|_{V,T}$ in which $A_i$ is the Helmholtz free energy of the $i$th cell) as the main building block of the transition rules of electrons between cells is obtained using the relation $A_{\tilde{N}}(V,T)-T\left(\frac{\partial A}{\partial T}\right)_{\tilde{N},V}=\left\langle E\right\rangle$. By considering the fact that $\mu(T\rightarrow 0)\rightarrow 0$ and $ \zeta_{\phi}(T)=aT^{-1/2}$ for two dimensional electron gas \cite{altshuler1985electron} ($a$ is a proportionality constant), one finds that (see APPENDIX A Ref.\cite{najafi2019electronic}):
\begin{equation}
\mu_i =k_BT\ln\left(e^{h_i}-1\right)+UT^{\frac{1}{2}}h_i-IZ_iT^{\frac{1}{2}}
\end{equation}
in which $U =\frac{2k_Bm\sqrt{2Da}e^2\pi^2}{8\epsilon_0\hbar^2}$, $I=\text{sinh}^{-1}(1)\frac{e^2}{\pi\epsilon_0\sqrt{Da}}$, $h_i=\frac{N_i}{T}$ and $i$ stands for the $i$th cell. The effect of randomness of $Z_i$'s (that are supposed to be random noise with an uniform probability measure), which captures the on-site (diagonal) disorder is investigated. The probability of adding a particle to the $i$th cell of the system is shown to be proportional to $\exp\left[-\beta \tilde{\mu}_i \right]$ in which $\tilde{\mu}_i\equiv \mu_i-\mu_0$ and $\mu_0$ is the average chemical potential of the system, whereas the probability of the transition between two sites (say cell 1 $\rightarrow$ cell 2) is obtained by 
\begin{equation}
\text{relative probability}=e^{-\beta\left(\mu_2-\mu_1\right)}
\label{relative0}
\end{equation}
for which the following relation is used:
\begin{equation}
\mu_2-\mu_1=k_BT\ln\left(\frac{e^{h_2}-1}{e^{h_1}-1}\right) +UT^{\frac{1}{2}}(h_2-h_1)-IT^{\frac{1}{2}} (Z_2-Z_1).
\end{equation}
these relations are of especial importance in the following sections. This is the base of the model proposed.
 Consider figure \ref{Honeycomba} in which an electron system has been divided into some hexagonal cells. We have shown a cell and its neighbors in fig.~\ref{Fig:Transport} (see also Fig.~\ref{Fig:Movement}) each of which has its own local chemical potential $\mu_i$, $i=0,1,2,...,6$. Let us consider for a moment that the order of potentials is $\mu_i<\mu_j$ for $i<j$. According to Eq. (A.2). Ref.~\cite{najafi2019electronic} the site $0$ is said to be \textit{unstable} (has the potential to give an electron to its neighbors) if its chemical potential $\mu(0)$ exceeds the chemical potential of the bulk $\mu_0$. If the mentioned site is unstable, it has the potential to release electrons to the neighbors, and the first candidate for this charge transfer is the neighbor with the smallest chemical potential $\mu$, i.e. $\mu_1$ here. After this process (whether the charge transfer to the first neighbor has taken place or not) the next candidate for the electron transfer is the site with the nearest $\mu$ to $\mu_1$, i.e. $\mu_2$, etc. The Metropolis Monte Carlo method~\cite{gould1996computer} is employed for these charge transfers, i.e. the electron transport from $0$ to any site $i$ is occurred with the probability:
\begin{equation}
P_{0\rightarrow i}\sim \left\lbrace \begin{array}{c} \Theta(\mu(0)-\mu_0) \times\text{Max}\left\lbrace 1,e^{-\beta\left( \mu(i)-\mu(0)\right)} \right\rbrace \\ 0 \end{array}\right.
\label{transport_prob}
\end{equation}
ّfor which the first one is for the case $0$ and $i$ are neighbors, and the second line is for the other cases, and $\Theta(x)$ is the step function. \\

Using this probability, Najafi showed that the system undergoes a percolation transition inline in the $T-\Delta$ phase space, realizing the observed metal-insulator transitions in 2D electron gases (2DEG)~\cite{kravchenko1994possible}. A separate study on the $1/f$ noise in 2DEG is currently being done based on the same physics explained above.

\section{conlusion}
In this paper, we reviewed the SOC concepts in various systems. First, we presented some examples, including the systems that show SOC, like earthquake, rain falling \textit{etc}. In the second part, we presented the evidence showing that the BTW sandpile model is $c=-2$ LCFT, and is tied to $W$-algebras. The simulation results for SOC in various systems were presented in the last part. There we considered the SOC in fluid propagation in porous media, in cumulus clouds, in an excitable random system, in imperfect supports, and in 2DEG. We also considered vibrating ASM, invasion sandpile model, and diffusive sandpiles.

\bibliography{refs}

\end{document}